\documentclass[%
twocolumn,
superscriptaddress,
 amsmath,amssymb,
 aps,
]{revtex4}

\usepackage{graphicx}
\usepackage{dcolumn}
\usepackage{bm}
\usepackage{todonotes}
\usepackage{longtable}
\usepackage[version=4]{mhchem}
\usepackage[hidelinks]{hyperref}
\usepackage{siunitx}

\usepackage{pdfpages}

\begin{document}

\title{First-principles screening of materials with extreme effective masses}

\author{Szymon Błazucki}
\affiliation{Theory of Condensed Matter, Cavendish Laboratory, University of Cambridge, Cambridge, United Kingdom}
\author{Junfeng Qiao}%
 \email{junfeng.qiao@epfl.ch}
\affiliation{Theory and Simulation of Materials (THEOS) and National Centre for Computational Design and Discovery of Novel Materials (MARVEL), \'Ecole Polytechnique F\'ed\'erale de Lausanne (EPFL), Lausanne, Switzerland}
\affiliation{
PSI Center for Scientific Computing, Theory and Data, Paul Scherrer Institute, %
Villigen PSI, Switzerland%
}%
\author{Aleksandr Poliukhin}
\affiliation{Theory and Simulation of Materials (THEOS) and National Centre for Computational Design and Discovery of Novel Materials (MARVEL), \'Ecole Polytechnique F\'ed\'erale de Lausanne (EPFL), Lausanne, Switzerland}
\author{Nicola Marzari}
\affiliation{Theory of Condensed Matter, Cavendish Laboratory, University of Cambridge, Cambridge, United Kingdom}
\affiliation{Theory and Simulation of Materials (THEOS) and National Centre for Computational Design and Discovery of Novel Materials (MARVEL), \'Ecole Polytechnique F\'ed\'erale de Lausanne (EPFL), Lausanne, Switzerland}
\affiliation{
PSI Center for Scientific Computing, Theory and Data, Paul Scherrer Institute, %
Villigen PSI, Switzerland%
}%

\date{\today}

\begin{abstract}
The effective mass of charge carriers is a fundamental descriptor of the electronic structure of materials, and can be used to assess performance in electronics applications, or to screen for thermoelectrics and transparent conductors. Here, we perform a high-throughput computational screening of approximately 20,000 experimentally known three-dimensional stoichiometric inorganics obtained from the Materials Cloud 3D structure database.
By combining density-functional theory calculations and maximally localized Wannier functions, we are able to compute the full conductivity effective mass tensor for electrons and holes from the Boltzmann transport equation in the constant relaxation-time approximation. This approach captures the effects of band non-parabolicity, anisotropy, and valley multiplicity that would be neglected by standard parabolic fittings. The screening identifies a curated set of candidates exhibiting extreme electronic properties, from ultra-low to ultra-large effective masses, these latter associated with flat-band physics. We validate the workflow by recovering established high-mobility semiconductors and highlight promising novel candidates. Furthermore, we classify materials by their mass anisotropy and discuss the physical limits of defining a conductivity effective mass in narrow-gap regimes at room temperature. Importantly, the resulting dataset provides a systematic roadmap to search for high-performance materials in novel chemical spaces.
\end{abstract}

\maketitle

\section{\label{sec:level1}Introduction}

The effective mass of charge carriers is a central concept in solid-state physics and materials science, serving as a bridge between electronic band structure and observable transport phenomena. It captures how electrons and holes respond to electric fields, thus entering into the determination of electrical conductivity, charge mobility, and thermoelectric efficiency \cite{ashcroft}. Materials with a low effective mass are particularly sought after in high-performance electronic and optoelectronic applications. These include transparent conducting oxides, where high mobility combined with optical transparency is required \cite{no_ceder, transparent_conducting}, as well as thermoelectric materials, where low effective mass enhances electrical conductivity and can contribute to a high power factor \cite{Suwardi2019InertialEvaluation, boltztrap_eff_mass}. More broadly it is also relevant for achieving low light yield in scintillators, where mobility affects recombination rates \cite{different_heuristics}.

Variations in effective mass, whether ultra-low, ultra-high, associated with flat and isolated bands, or strongly anisotropic, play a central role in shaping diverse classes of electronic behaviour. In thermoelectrics, different flavours of effective mass, including inertial, Seebeck, and conductivity effective masses \cite{boltztrap_eff_mass, Pei2011ConvergenceThermoelectrics}, capture complementary aspects of electronic transport and energy conversion. Optimizing these quantities is critical for enhancing the thermoelectric figure of merit ($ZT$) \cite{Suwardi2019InertialEvaluation, mahanBestThermoelectric1996}, and many state-of-the-art thermoelectric compounds exhibit relatively low conductivity effective masses that promote efficient charge transport while maintaining favourable Seebeck coefficients \cite{Pei2011ConvergenceThermoelectrics}. At the opposite extreme, materials with a high effective mass often display suppressed mobility, enhanced carrier localization, or strong electron-phonon coupling, characteristics relevant to thermal insulators, energy storage materials, and systems with polaronic behaviour \cite{Franchini2021PolaronsMaterials}. Large effective masses can also arise from electronic flat bands and are commonly associated with correlated-electrons phenomena \cite{Cao2018UnconventionalSuperlattices, Regnault2022CatalogueMaterials}, including magnetism, Wigner crystallization, and unconventional superconductivity in systems such as magic-angle twisted bilayer graphene \cite{flat_band}. To preserve these characteristics, flat bands must typically be well isolated from neighbouring bands to minimize hybridization \cite{Lu2020LithiumBands, Regnault2022CatalogueMaterials}. Beyond such extremes, pronounced anisotropy in the effective mass, where its value varies strongly with direction in $k$-space, leads to direction-dependent transport and optical responses, particularly in low-symmetry crystals, layered compounds, and van-der-Waals heterostructures \cite{Novoselov20162DHeterostructures}. Together, these regimes underscore the broad physical relevance of effective masses as descriptor for identifying materials with unusual or extreme electronic behaviour of great relevance for industrial applications.

Given the wide range of physical phenomena associated with extreme or anisotropic effective masses, discovering such materials requires approaches that can efficiently navigate large chemical and structural spaces. High-throughput materials screening has become a core strategy in computational materials discovery, enabling the exploration of such spaces with first-principles accuracy \cite{Curtarolo2013}. The combination of automated workflows, materials databases, and machine learning algorithms accelerates the search for compounds with desirable properties such as high conductivity, large Seebeck coefficients, or structural stability \cite{Pizzi2016AiiDA:Science, Jain2013Commentary:Innovation, Oganov2019StructureDiscovery, Liu2017MaterialsLearning}. It facilitates both the identification of candidate materials for targeted applications and the discovery of unexpected compounds with novel functionalities. Within this framework, the effective mass remains a powerful tool in high-throughput screening. It is computationally inexpensive to calculate, interpretable in physical terms, and applicable across a wide variety of material classes; thus well suited in identifying candidates for more detailed and expensive investigation. Despite its conceptual simplicity, several recent works have successfully applied effective-mass-based heuristics to identify new functional materials, including transparent conductors, thermoelectrics, and flat-band compounds \cite{Wang2022High-ThroughputSemiconductors, Li2023High-throughputLimit, Zhang2020FlatDichalcogenides}. However, systematic high-throughput evaluations of effective masses across large sets of experimentally realized 3D crystal structures remain limited.

In this work, we perform a large-scale screening of around \num{20000} three-dimensional crystal structures from the Materials Cloud 3D structure database (MC3D) \cite{HuberSebastiaan2022MaterialsMC3D,huber_mc3d_2025}, computing their conductivity effective mass from first-principles electronic structure calculations. Importantly, MC3D entries are all obtained from experimentally known stoichiometric inorganics reported in ICSD, MPDS \& COD \cite{Zagorac2019RecentFeatures,Blokhin2018, Grazulis2012}. The present aim is to identify materials with extreme effective masses — either ultralow, ultrahigh, or highly anisotropic — across a chemically diverse set of experimentally known compounds. This is enabled by recent advances in computational infrastructure, automation frameworks, and open materials databases, which allow for robust and scalable evaluation of electronic properties \cite{Pizzi2016AiiDA:Science, Zagorac2019RecentFeatures}. By focusing on the effective mass as a physically grounded, computationally accessible descriptor, our work contributes to this growing body of high-throughput materials design and provides a valuable dataset for future studies on charge transport, band structure topology, and anisotropic behaviour in crystalline solids.

\section{\label{sec:results}Results}

\begin{figure*}[!htb]
\includegraphics[width=0.95\textwidth]{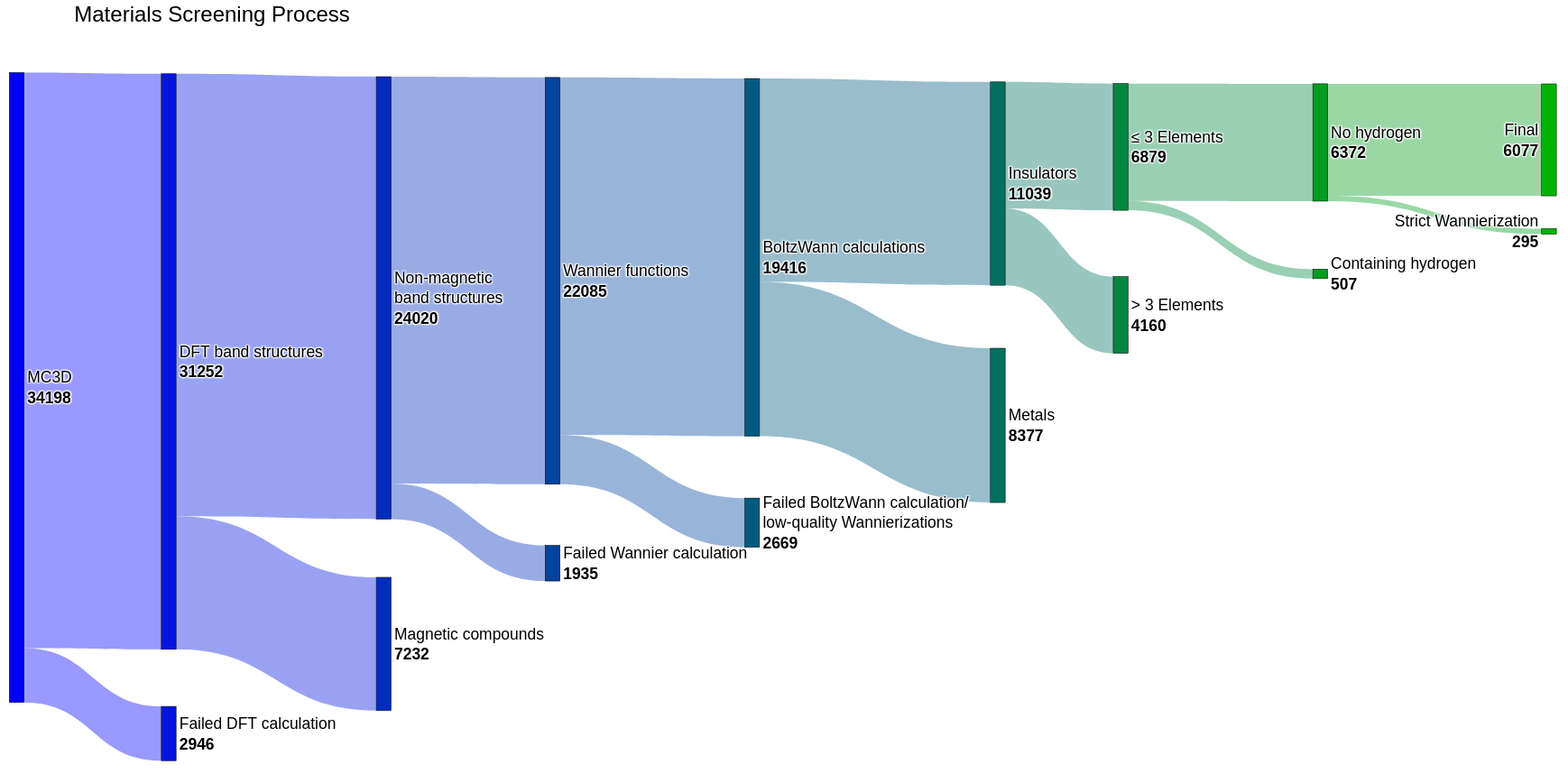}
\caption{\label{fig:sankey}Filtering process diagram presenting how set of materials used in this study were obtained from the MC3D dataset. Failed entries correspond mostly to exceeded-memory-limit errors, and less frequently to compute-node errors or code crashes.}
\end{figure*}
\subsection{Screening and Filtering Criteria}

The first stage of the workflow consists of filtering the \num{34198} structures in the MC3D database \cite{HuberSebastiaan2022MaterialsMC3D} to retain only systems suitable for effective-mass analysis. Figure \ref{fig:sankey} summarizes the full screening pipeline. At the very beginning, only non-magnetic materials are selected. Then, structures were discarded if the construction of maximally localized Wannier functions or the subsequent semiclassical Boltzmann transport calculations (using BoltzWann) failed, forming the starting dataset for this work. Afterwards, we begin by removing metallic systems, identified as those with Kohn-Sham band gaps at the PBEsol level smaller than 0.04 eV, since a meaningful conductivity effective mass cannot be defined in the absence of a band edge. This step reduces the dataset to \num{11039} insulating compounds.

To simplify the subsequent analysis and limit chemical complexity, we further restrict the dataset to materials containing at most three distinct chemical elements. This criterion, although strict, excludes only a 4160, leaving \num{6879} compounds. Among these, materials containing hydrogen must be treated with caution: hydrogen positions are often poorly resolved in experimental crystal structures, leading to structural uncertainties and unreliable DFT structural optimization. We therefore discard H-containing compounds, resulting in a high-quality dataset of \num{6372} structures ready for transport analysis (green branch in Fig. \ref{fig:sankey}).

\subsection{Computation of Conductivity Effective Masses}

Effective masses in high-throughput studies are typically estimated along high-symmetry paths in the Brillouin zone \cite{different_heuristics, transparent_conducting}. While such 1D sampling provides a rough order-of-magnitude estimate, it neglects the full k-space curvature and is unable to capture band anisotropy, valley multiplicity, or nonparabolicity. As a consequence, the resulting effective masses can be significantly biased, especially in systems with multiple near-degenerate extrema or flat dispersions. 

One of the earliest dense-sampling approaches was introduced in Ref. \cite{boltztrap_eff_mass}, where BoltzTraP calculations \cite{Madsen2006BoltzTraP.Quantities} were used to interpolate band energies on uniform grids and to compute semi-classical transport coefficients in the constant-relaxation-time approximation.
The method presented here adopts a similar screening approach to \cite{boltztrap_eff_mass}, but instead we leverage maximally localised Wannier functions (MLWFs) \cite{wannier_review} to accurately interpolate band energies and velocities on a densely sampled Brillouin zone for obtaining converged conductivity.
Compared with the Fourier-based interpolation used in BoltzTraP, the Wannier-based approach is especially accurate at resolving band crossings, which can have a large influence on transport phenomena, such as the nonlinear response induced by Weyl points \cite{Du2021NonlinearEffects}; besides, the velocity matrix elements are computed analytically from the Wannier Hamiltonian, achieving higher accuracy than the finite-difference approach.
Calculations are performed using the BoltzWann code \cite{boltzwan} which implements the same equations as Ref. \cite{Madsen2006BoltzTraP.Quantities} but using Wannier interpolations. The obtained full conductivity tensor, and thus the complete effective mass tensor, can be compared against experimental data.
It is important to mention here that to ensure high interpolation quality, only Wannierizations with a band distance \cite{bands_distance_fermi} smaller than \qty{30}{meV} were retained. This is the final step of the filtering diagram in Fig.~\ref{fig:sankey} where a more strict criterion on the quality of Wannierization was enforced, resulting in the final set of \num{6077} materials. The band distance is a measure of the quality of band Wannierization, closely resembling the root mean square error between the plane wave (PW) and the Wannierized bands \cite{bands_distance_fermi,bands_distance}. Calculations failing this criterion appear in Fig. \ref{fig:sankey} under ‘Strict Wannierizations’.

Following the approach of Ref.~\cite{boltztrap_eff_mass} for the calculation of conductivity effective masses the inverse effective mass tensor is calculated from the conductivity tensor $\sigma_{\alpha \beta}(\mu, T)$ as:
\begin{equation}\label{eqn:boltztrap}
\left(m_{\mathrm{c}_{\alpha \beta}}^*(\mu, T)\right)^{-1}=\frac{\sigma_{\alpha \beta}(\mu, T)}{e^2 \tau} \times \frac{1}{n(\mu, T)},
\end{equation}
where $\tau$ is the relaxation time (for which the common value of \qty{10}{fs} is used), $e$ is the electron charge, and $n(\mu, T)$ is the net charge carrier concentration expressed as a function of chemical potential $\mu$ and temperature $T$. 
Although using $\mu$ is convenient, as it is part of the default output of BoltzWann, it is often more intuitive to interpret conductivity in terms of doping levels $n$. The connection between the two can be written as
\begin{equation}\label{eqn:electrons}
    n  + \frac{N_{\text {val }}}{V} = \int g(E) f(E; \mu, T) dE, 
\end{equation}
where $N_{\text{val}}$ is the number of valence electrons, satisfying the relation
\begin{equation}
    \frac{N_{\text {val }}}{V} = \int g(E) f(E; \mu=E_{\textrm{F}}, T=0) dE, 
\end{equation}
where $f(E; \mu, T)$ is a Fermi-Dirac distribution at chemical potential $\mu$ and temperature $T$, and $g(E)$ stands for the density of states, which is an optional output of the BoltzWann code. For a given temperature Eq. \ref{eqn:electrons} can be easily solved numerically for $\mu$, and the effective mass tensor in Eq. \ref{eqn:boltztrap} can be treated as a function of doping $n$ and temperature $T$ only. More details of the above equations can be found in the Supplementary Material. 

\subsection{Effective mass distribution}

\begin{figure}[!htb]
\includegraphics[width=0.45\textwidth]{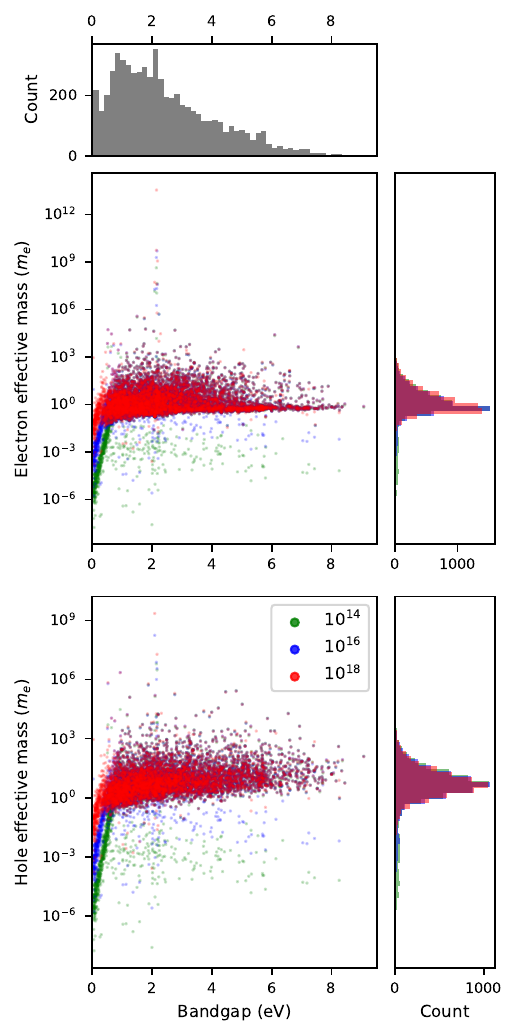}
\caption{\label{fig:effmass_vs_bandgap}Maximum absolute eigenvalue of the conductivity effective-mass tensor as a function of the Kohn-Sham PBEsol band gap, at $T=\qty{300}{K}$ and doping levels of $10^{14}$, $10^{16}$, and $10^{18}~\text{carriers/cm}^{-3}$, for the full set of semiconductors considered in this work. The diagonal features at small gaps arise from intrinsic thermally excited carriers (see text).}
\end{figure}

Figure~\ref{fig:effmass_vs_bandgap} summarizes the dependence of the maximum absolute effective mass tensor eigenvalue on the band gap at three different doping levels for the full set of semiconductors considered in this work. We focus on eigenvalues, because the principal components of the effective-mass tensor provide an orientation-independent measure of band curvature. Materials with small band gaps exhibit a particularly wide spread of effective masses, spanning several orders of magnitude. Points concentrated near zero band gap correspond to compounds that are nearly Kohn-Sham metals at the PBEsol level and are affected by the non-zero temperature broadening of the Fermi-Dirac distribution. Because Kohn-Sham DFT at the semi-local level underestimates band gaps, many of these very small-gap entries are likely to be semiconductors with more moderate transport behaviour when described by calculations involving higher levels of theory. The impact of temperature and band-gap underestimation on these trends is analysed in detail in the next Section.

\begin{figure}[!htb]
\includegraphics[width=0.5\textwidth]{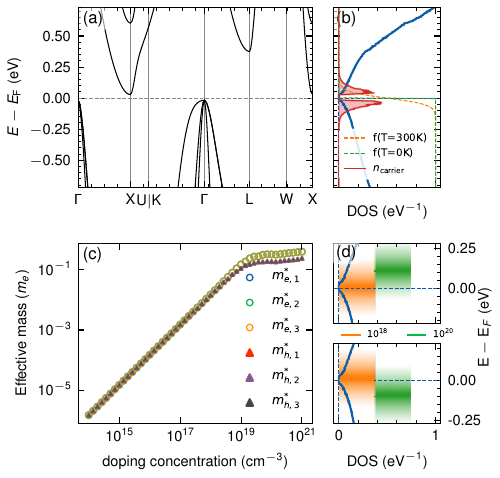}
\caption{\label{fig:flatband}Effect of doping on effective mass. a) calculated band structure for small band gap \ce{GaLiSi}, b) corresponding density of states with Fermi-Dirac distribution centred at Fermi level at 0K and 300K. The red line is $n_{carrier}$ which is a product of the density of states and Fermi distribution $f(T=300K)$ for holes at the valence band side, and electrons at the conduction band side. c) calculated effective masses in principal directions for electrons and holes as a function of doping. d) enlarged valence band maximum (VBM) and conduction band minimum (CBM) with shaded rectangles indicating range of Fermi Dirac functions where $10^{-3}<f(T=300K)<1-10^{-3}$. It can be seen (see text) that only above a doping concentration of $10^{19}\text{carriers/cm}^3$ the effect of intrinsic carriers is washed out.}
\end{figure}

The final step before preparing a ranked list of candidates is to select an appropriate doping level. The key to capture the physics is to probe only the edge of the conduction or valence band. As discussed in \cite{ashcroft}, these are the only electrons or holes that contribute to electrical conductivity, so the chemical potential should be positioned close to the band edges. This approach provides an estimate of the conductivity effective mass, which is relatively easy to measure and can be thought of as an average of all the peaks close to the valence or conduction band edges. We found the value of $10^{18}~\text{carriers/cm}^{-3}$ to give the best results, consistent with prior studies \cite{boltztrap_eff_mass}.

To illustrate the connection between band dispersions and the resulting conductivity effective masses, we provide an illustrative example for \ce{GaLiSi} in Figure \ref{fig:flatband}. It shows the calculated chemical potentials at \qty{300}{K} for doping levels of $10^{18}$ and $10^{20}~\text{carriers/cm}^{-3}$ together with the range where the Fermi-Dirac distribution at these doping levels differs from its extreme values (0 and 1) by a threshold of $10^{-3}$, overlaid on the density of states (DOS). As can be observed, dopings in this range do not yield similar results, and the effective mass value only stabilizes above $10^{19}~\text{carriers/cm}^{-3}$.
This doping dependence follows directly from Eq.~\ref{eqn:boltztrap}, in which $\left(m^*_{\mathrm{c}}\right)^{-1} \propto \sigma / n$. At low doping the conductivity of a narrow-gap material is dominated by intrinsically (thermally) excited carriers, whose concentration is fixed by the gap and the temperature; $\sigma$ is therefore almost insensitive to the added doping, and the apparent effective mass grows roughly linearly with $n$. Once the doping exceeds the intrinsic carrier concentration, extrinsic carriers dominate, $\sigma$ scales with $n$, and the effective mass saturates at its physical band value — precisely the behaviour seen in Figure \ref{fig:flatband}c, and the origin of the order-of-magnitude difference between the values extracted at $10^{18}$ and $10^{20}~\text{carriers/cm}^{-3}$
 for the small-gap cases.
The electrons and holes present in the calculation predominantly come from the band edges. Although the rectangles in Figure \ref{fig:flatband}d might appear wide due to the threshold ($10^{-3}$) and temperature ($300 \: K$) used, these doping levels are just above the band edges. 
The slow increase in effective masses with doping reflects the gradual sampling of nonparabolic portions of the bands, while the much steeper, near-linear rise observed for small-band-gap materials in Fig. \ref{fig:flatband}c is the intrinsic-carrier effect described above. 
Additional plots for lower temperatures and discussion on doping sensitivity for different materials are available in the Supplementary Material.


\section{Discussion}

\subsection{Candidate validation and dataset reliability}

After selecting a representative doping level and sufficient bandgap, we compiled lists of promising candidates for low and high electron/hole effective masses, anisotropic effective masses, and flat, isolated electronic bands.
The most notable low-effective-mass compounds, grouped by structural prototype, are summarised in Table \ref{tab:combined_materials}.
Complete lists for all categories are provided in the Supplementary Material.

\begin{table*}[ht!]
\centering
\caption{\label{tab:combined_materials}Overview of effective mass candidates grouped by electronic character and mass category. Materials are clustered by their chemical prototype, obtained by the isovalent-substitution mapping described in the Methods section; a label such as \ce{CuBO2} therefore denotes a whole isovalent family and not a specific compound. Multiple entries of the same formula mean that polymorphs of the same material were found. In Parts II and III the high-mass and strongly anisotropic rankings were found to overlap almost completely, so rather than reproducing two near-identical tables we report them as a single merged list of candidates that are both high-mass and anisotropic. For the full list with ranking positions please refer to the Supplementary Material.}
\begin{tabular}{l l} 
\hline \hline 
\textbf{Prototype} & \textbf{Materials} \\
\hline
\multicolumn{2}{c}{} \\[-2ex] 
\rlap{\textit{\textbf{Part I: Low effective mass candidates}}} & \\
\hline
\ce{CoC} & \parbox[t]{11cm}{\raggedright \ce{SiRh}, \ce{FeSi}, \ce{GeRu}, \ce{SiRu}} \\
\ce{CuBO2} & \parbox[t]{11cm}{\raggedright \ce{InAgS2}, \ce{InAgTe2}, \ce{GaCuSe2}, \ce{GaCuTe2}, \ce{GaAgSe2}, \ce{AlCuSe2}, \ce{AlAgTe2}, \ce{AlCuTe2}, \ce{GaCuS2}, \ce{AlAgSe2}, \ce{GaAgS2}} \\
IV--VI & \parbox[t]{11cm}{\raggedright \ce{SnS}, \ce{GeSe}, \ce{SnTe}, \ce{GeTe}, \ce{PbS}, \ce{TePb}} \\
\ce{As}-\ce{Te} & \parbox[t]{11cm}{\raggedright \ce{Ge(Te2As)2}, \ce{Te3As2}, \ce{AgBiSe2}, \ce{Bi2Te2S}, \ce{Sb2TeSe2}, \ce{Sb2Te4Pb}} \\
\ce{Be2C} & \parbox[t]{11cm}{\raggedright \ce{Mg2Ge}, \ce{Mg2Si}} \\
III-V & \parbox[t]{11cm}{\raggedright \ce{GaAs}, \ce{InP}, \ce{GaN}, \ce{GaN}} \\
\ce{ZnCN2} & \parbox[t]{11cm}{\raggedright \ce{ZnGeAs2}, \ce{CdSnP2}, \ce{CdSiAs2}, \ce{CdGeP2}} \\
\ce{HBeN} & \parbox[t]{11cm}{\raggedright \ce{LiMgBi}, \ce{Sr(MgBi)2}, \ce{Ba(MgBi)2}, \ce{Mg3Sb2}, \ce{Ca(MgBi)2}, \ce{NaMgAs}} \\
\ce{ZnO} & \parbox[t]{11cm}{\raggedright \ce{CdSe}, \ce{CdTe}, \ce{CdSe}, \ce{CdTe}, \ce{ZnTe}, \ce{ZnTe}, \ce{CdS}, \ce{ZnSe}, \ce{CdS}, \ce{ZnO}, \ce{ZnS}, \ce{ZnS}, \ce{ZnS}} \\
\ce{HZnN} & \parbox[t]{11cm}{\raggedright \ce{Li2AgSb}, \ce{LiZnAs}, \ce{LiCdP}, \ce{NaZnAs}, \ce{LiZnN}, \ce{KCdAs}} \\
\ce{Zn2CO4} & \parbox[t]{11cm}{\raggedright \ce{Sn(HgSe2)2}, \ce{Hg2GeSe4}} \\
\ce{InSe} & \parbox[t]{11cm}{\raggedright \ce{GaTe}, \ce{TlInS2}, \ce{GaSe}, \ce{GaSe}, \ce{InSe}, \ce{GaSe}, \ce{TlInTe2}, \ce{InSe}} \\
\ce{ScNiN} & \parbox[t]{11cm}{\raggedright \ce{YSbPt}, \ce{ScNiBi}, \ce{ScBiPd}} \\
\ce{H3N} & \parbox[t]{11cm}{\raggedright \ce{Na3P}, \ce{Na3Sb}, \ce{KNa2Sb}, \ce{Cs3Bi}, \ce{CsK2Sb}, \ce{K3P}, \ce{K3Sb}, \ce{Cs3Sb}} \\
\ce{ZnB2O4} & \parbox[t]{11cm}{\raggedright \ce{Cd(InTe2)2}, \ce{In2HgTe4}, \ce{Cd(InSe2)2}, \ce{Zn(InS2)2}, \ce{In2HgSe4}, \ce{Cd(InSe2)2}, \ce{Cd(InSe2)2}, \ce{Cd(InTe2)2}, \ce{Zn(InSe2)2}, \ce{Zn(InTe2)2}, \ce{Al2HgTe4}, \ce{Zn(InSe2)2}, \ce{Zn(GaSe2)2}, \ce{Ga2HgTe4}, \ce{Ga2HgSe4}, \ce{Cd(GaSe2)2}} \\
\ce{H2HN} & \parbox[t]{11cm}{\raggedright \ce{K2NaAs}, \ce{Na2LiN}} \\
\ce{BeCO3} & \parbox[t]{11cm}{\raggedright \ce{BaSnO3}, \ce{CaSnO3}} \\
\ce{Be3NN} & \parbox[t]{11cm}{\raggedright \ce{Sr3SbN}, \ce{Mg3SbN}} \\
Others & \parbox[t]{11cm}{\raggedright \ce{AgO}, \ce{SiB6}, \ce{Li2MgSi}, \ce{TiFe2Ge}, \ce{ZrSe2}, \ce{YN}, \ce{BaSiC}, \ce{Be5Pt}, \ce{Ba2NF}, \ce{FeTe2}, \ce{VCu3O4}, \ce{Al2Ru}, \ce{GaReSi}, \ce{MgGeAs2}, \ce{CuBr}, \ce{NaCuTe}, \ce{GaGeTe}, \ce{SiPAu}, \ce{NbCuN2}, \ce{MgTe}, \ce{TlBiS2}, \ce{Ca(CuS)2}, \ce{CdIBr}} \\
\hline
\multicolumn{2}{c}{} \\[-2ex]
\rlap{\textit{\textbf{Part II: High and anisotropic electron effective mass candidates}}} & \\
\hline
\ce{TiFN} & \parbox[t]{11cm}{\raggedright \ce{TiNCl}, \ce{ZrNCl}, \ce{ZrBrN}, \ce{TiBrN}, \ce{HfBrN}} \\
\ce{BeTiN2}  & \parbox[t]{11cm}{\raggedright \ce{CaTiN2}, \ce{SrTiN2}, \ce{BaHfN2}} \\
\ce{ScOF} & \parbox[t]{11cm}{\raggedright \ce{ScBrO}, \ce{YClO}} \\
IV--VI & \parbox[t]{11cm}{\raggedright \ce{SnS}} \\
\ce{TiCO3} & \parbox[t]{11cm}{\raggedright \ce{TiPbO3}} \\
\hline
\multicolumn{2}{c}{} \\[-2ex]
\rlap{\textit{\textbf{Part III: High and anisotropic hole effective mass candidates}}} & \\
\hline
Ruddlesden-Popper & \parbox[t]{11cm}{\raggedright \ce{Rb2ZnF4}, \ce{Rb2HgF4}} \\
Layered MNX & \parbox[t]{11cm}{\raggedright \ce{TiNCl}, \ce{TiBrN}} \\
IV--VI & \parbox[t]{11cm}{\raggedright \ce{SnS}} \\
\ce{Zn} & \parbox[t]{11cm}{\raggedright \ce{Zn}, \ce{Cd}} \\
Thio-spinels & \parbox[t]{11cm}{\raggedright \ce{Mg(AlS2)2}} \\
\ce{ScF3} & \parbox[t]{11cm}{\raggedright \ce{ScF3}} \\
\hline \hline
\end{tabular}
\end{table*}

\begin{figure*}[!htb]
\includegraphics[width=0.99\textwidth]{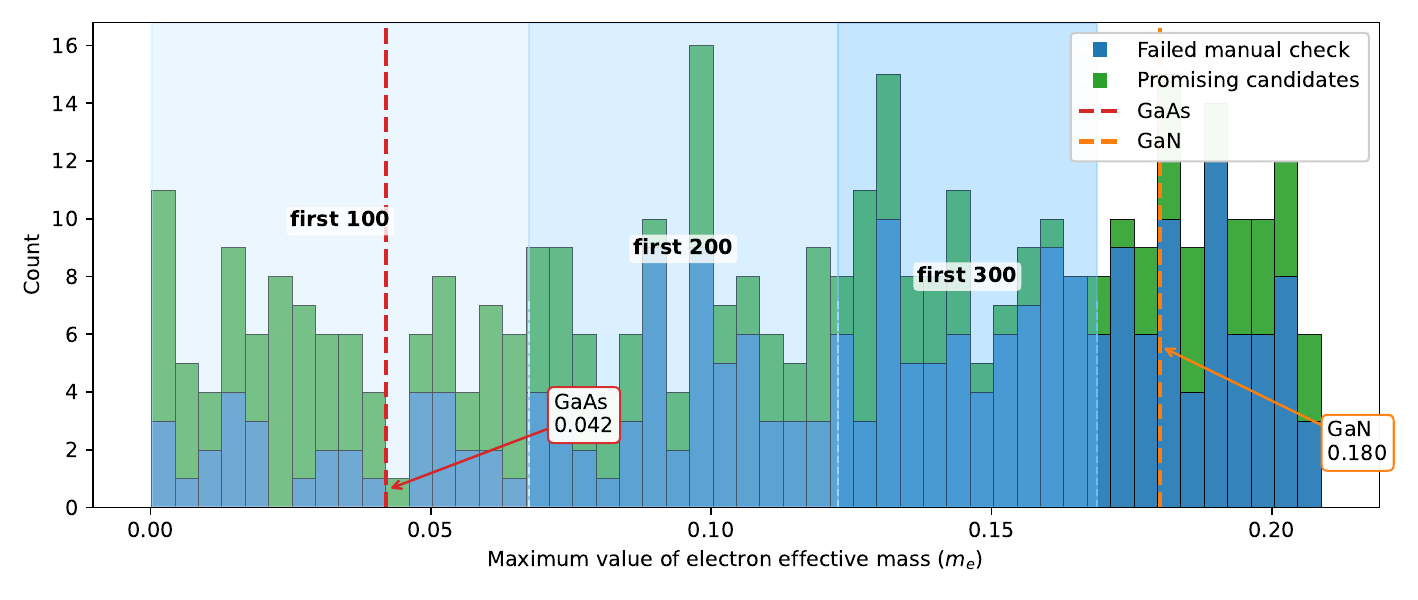}
\caption{\label{fig:histogram}Histogram of the electron effective mass for the top low-mass candidates at a doping of $10^{18}~\text{carriers/cm}^{-3}$, with columns coloured by the outcome of the manual validation against the experimental literature.}
\end{figure*}

Importantly, to verify that automatically screened entries correspond to experimentally feasible materials, we manually inspected the original crystallographic references for all shortlisted candidates. This resulted in checking around 900 publications to confirm phase stability under ambient conditions, toxicity, stoichiometry, experimental unit cell volume compared to the one obtained after structural relaxation, and reactivity with air humidity.
Figure \ref{fig:histogram} illustrates the outcome of this validation for low electron effective mass at a doping of $10^{18}~\text{carriers/cm}^{-3}$.
Green bars denote compounds that passed manual verification, whereas blue bars correspond to structures rejected due to unreliable or incomplete experimental data.
A substantial fraction of the lowest-mass candidates were excluded during this step, highlighting the necessity of careful a posteriori quality control in high-throughput studies. Because our aim was to ensure that we did not miss any potentially promising candidates, the aforementioned manual validation was performed on the top 400 low electron-mass entries and 350 low hole-mass entries.
For the remaining categories—high effective mass, anisotropy, and flat bands—the top 100 candidates were inspected. This choice ensured that well-known high-mobility compounds (e.g. GaAs, GaN, ZnO) fall within the inspected window, while also capturing less-explored materials with comparable or superior performance.

The grouped list of low and high effective mass materials with primitive unit cells containing a maximum 8 atoms is presented in Table \ref{tab:combined_materials}. It represents one of the key outcomes of this work. Highly anisotropic effective mass materials, i.e., ones with the highest standard deviation of 3 tensor eigenvalues, were checked as well, but due to very strong overlap with high effective mass candidates, tables for these were merged. To provide a better understanding of these results, we organised candidates into chemically and structurally meaningful prototype families. This grouping highlights recurring motifs associated with high carrier mobility, such as chalcopyrite-derived phases (\ce{CuBO2}-type), IV-VI chalcogenides, III-V semiconductors, and several pnictide and bismuthide families. The presence of both well-established high-mobility materials (e.g., \ce{GaAs}, \ce{InP}, \ce{ZnO}) and less-explored compounds with comparable predicted performance indicates that the workflow captures known trends while revealing new and chemically diverse candidates. By providing prototype families rather than only individual formulas, the table offers a guidance for future experimental synthesis and computational exploration.

Despite very promising results in the table, this method has to be confronted with other approaches. Candidates lists were compared against other screening attempts performed in the literature and found to match them well. Although the values reported in Ref. \cite{transparent_conducting} based on high-symmetry-path calculations could  initially appear inconsistent with our results, a closer examination of their Supplementary Information \cite{transparent_conducting_appendix} shows that full Brillouin-zone sampling yields effective masses that are in very good agreement with those obtained here. This confirms reproducibility of the results.

For a quantitative benchmark, we compared the presented dataset against the ab initio electronic transport database of Ricci \textit{et al.} \cite{Ricci2017TransportDatabase}, which reports conductivity effective-mass tensors for approximately \num{48000} Materials Project entries, obtained from BoltzTraP Fourier interpolation of GGA(+U) band structures at the same doping level ($10^{18}~\text{carriers/cm}^{-3}$) and temperature (\qty{300}{K}) employed here. Matching entries between the two datasets by space group, reduced formula, number of sites, and unit-cell volume yields 1168 semiconductors in common. 
Supplementary material contains Figure which compares the sorted absolute eigenvalues of the effective-mass tensors: 93\% (90\%) of the electron (hole) eigenvalues agree within a factor of two, with a median relative deviation of about 11\% (18\%), despite the two workflows differing in interpolation scheme (Wannier vs.\ Fourier), exchange-correlation functional (PBEsol vs.\ PBE(+U)), pseudopotentials, and relaxed geometries. The population of larger discrepancies is dominated by narrow-gap compounds (e.g. \ce{GeTe}, \ce{SnTe}, \ce{GeSe}, and As--Ge--Te phases with gaps below \qty{0.1}{eV}), where, as discussed below, intrinsic thermal carriers make the conductivity effective mass ill-conditioned in both approaches; indeed, the vast majority of the outlying entries had already been flagged and rejected by our manual validation. This cross-database agreement supports the reliability of the Wannier-based workflow across chemically diverse compounds.


The screening also recovers a substantial number of materials with established or emerging technological relevance, which serves both as an external validation of the workflow and as a guide to where the novel candidates may find applications. Among the low-effective-mass candidates, \ce{GaAs}, \ce{InP}, and \ce{GaN} are commercially dominant platforms for high-frequency electronics, fibre-optic photonics, and power devices respectively; \ce{CdTe} is the leading thin-film photovoltaic technology; \ce{ZnO} is a widely deployed transparent conducting oxide; \ce{PbTe} is a workhorse of thermoelectric power generation \cite{Pei2011ConvergenceThermoelectrics}; and \ce{GeTe} underpins phase-change memory \cite{Wuttig2007PhaseChange}; \ce{BaSnO3} exhibits the highest room-temperature mobility among wide-gap perovskite oxides \cite{Kim2012BaSnO3} and is used as an electron-transport layer in perovskite solar cells \cite{Shin2017BaSnO3}; and the alkali antimonides (\ce{Cs3Sb}, \ce{CsK2Sb}, \ce{KNa2Sb}) are standard high-quantum-efficiency photocathodes in photomultipliers and accelerator photoinjectors \cite{Dowell2010Photocathodes, Mamun2015CsK2Sb}. Equally encouraging is the presence of compounds at the core of current research interest: \ce{Mg3Sb2} is among the most actively studied earth-abundant \textit{n}-type thermoelectrics \cite{Tamaki2016Mg3Sb2, Zhang2017Mg3Sb2}, with the isostructural Zintl $A$\ce{Mg2Bi2} ($A$ = Ca, Sr, Ba) phases attracting comparable interest as \textit{p}-type analogues \cite{Shuai2016CaMg2Bi2}, \ce{Mg2Si} is an established environmentally friendly mid-temperature thermoelectric \cite{Zaitsev2006Mg2Si}, \ce{TlInTe2} displays ultralow rattler-induced lattice thermal conductivity \cite{Jana2017TlInTe2, Pal2021TlInTe2}, \ce{InSe} is a high-mobility two-dimensional semiconductor \cite{Bandurin2017InSe}, the layered nitride halides \ce{TiNCl} and \ce{ZrNCl} found in our anisotropic high-mass lists become superconducting upon electron doping \cite{Yamanaka2010MNX}, and \ce{ScF3} is a prototype of pronounced negative thermal expansion \cite{Greve2010ScF3}. Beyond these, the screening highlights less-explored compounds, for instance \ce{Cu2Ga2Te4} (mpds S1210978), investigated as a thin-film photovoltaic absorber, and \ce{CdLiSb} (icsd 670568), which has been synthesised and characterised as a thermoelectric \cite{Yang2022LiCdSb}. A material-by-material summary of experimentally confirmed applications, compiled from the primary literature for all table entries, is provided in the Supplementary Material.

As an additional analysis, we created a list of flatband materials based on the band structure. For this purpose, we quantified the width of each ``band manifold'', defined here as a cluster of overlapping or touching bands that span a continuous energy interval, are separated by a finite energy gap from every other such cluster, and collectively contribute to the density of states. Materials whose highest occupied manifold has a narrow bandwidth, in this case below \qty{0.1}{eV}, must have a large hole effective mass due to their flat curvature. The energy separation between this narrow manifold and the next lower-energy manifold serves as a measure of band isolation, helping to identify separated flat band materials which might be of interest for certain applications. In the results of effective mass calculations for holes performed in this work, a significant overlap was observed between the set of materials with a high effective mass and materials with a narrow valence-band manifold, which confirms the applicability of this method for the search of these materials.

\subsection{Trends across the database}
Figure \ref{fig:effmass_vs_bandgap} shows the maximum principal effective mass as a function of band gap for dopings of $10^{14}$, $10^{16}$ and $10^{18}~\text{carriers/cm}^{-3}$.
Small-gap semiconductors exhibit the broadest spread in effective masses, spanning several orders of magnitude; while wide-gap insulators occupy a much narrower range.
A particularly dense cluster of points appears around band gaps of 2.0-2.3 eV and hole masses near 4~$m_e$, containing almost 60 different \ce{CdI2} polymorphs.

The diagonal linear groups of points near zero band gap originate from thermal broadening of the  Fermi-Dirac distribution at $T=300$K.
When the band gap becomes comparable to the thermal window sampled by the Fermi-Dirac distribution, both the valence and conduction band extremum contribute to transport, yielding incorrect effective masses.
Lowering the temperature narrows this sampling window and suppresses this effect.
As expected, repeating the analysis at $T=100$K (see Supplementary Material) significantly reduces the extent of these linear tails. However, the list of top candidates at both temperatures is similar, the temperature effect affecting mostly the value of effective mass, not the ordering. Because semilocal functionals such as PBEsol systematically underestimate band gaps \cite{Perdew2007RestoringSurfaces}, the linear dependency is expected to be diminished with more accurate electronic-structure methods (e.g. hybrid functionals or many-body approaches) for better band gap.

\begin{figure}
    \centering
    \includegraphics[width=\linewidth]{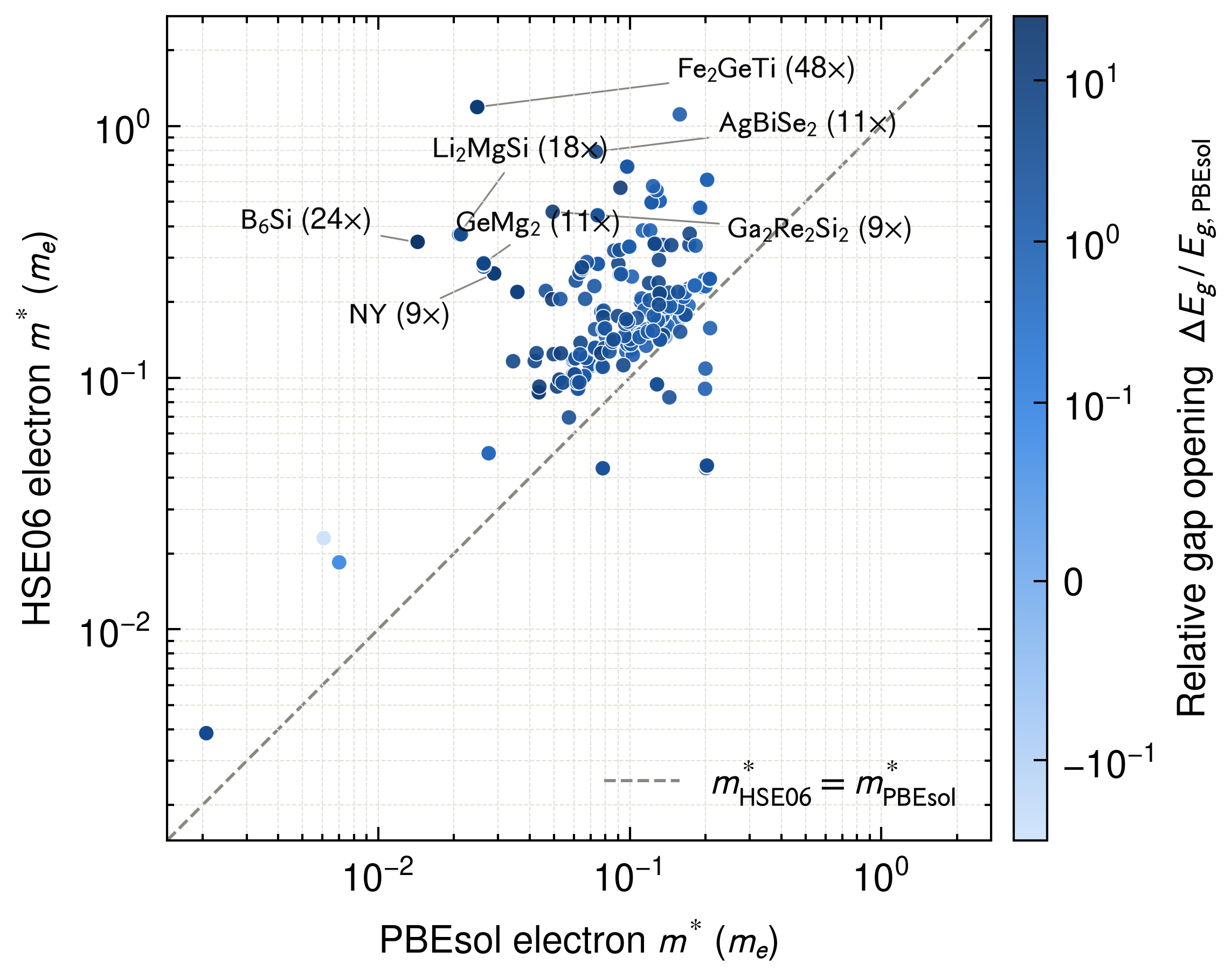}
    \caption{\label{fig:electron_parity}Parity plot of the electron conductivity effective mass obtained with the HSE06 hybrid functional against the PBEsol value, for the candidates of Table~\ref{tab:combined_materials} at a doping of $10^{18}~\text{carriers/cm}^{-3}$. The dashed line marks $m^{*}_{\mathrm{HSE06}} = m^{*}_{\mathrm{PBEsol}}$, so that points lying above it are compounds whose effective mass increases once the band gap is opened. Markers are coloured by the relative gap opening $\Delta E_g / E_{g,\mathrm{PBEsol}}$ on a symmetric-logarithmic scale. The ten largest outliers are annotated with the chemical formula followed, in parentheses, by the factor by which the HSE06 and PBEsol effective masses differ. The corresponding plot for holes is provided in the Supplementary Material.}
\end{figure}

To quantify this effect, we recomputed the band gap estimation and conductivity effective masses of 83 candidates from Table~\ref{tab:combined_materials} using the HSE06 hybrid functional \cite{Heyd2003HybridFunctionals, Krukau2006Screening}; by band gap estimation we mean the band gap obtained from the Wannier interpolation of the HSE06 band structure, which is more computationally manageable and accurate than direct HSE06 calculation of band structures. The physical origin of the spurious masses lies in the intrinsic carrier concentration, which scales as $n_i\propto (m^*_e m^*_h)^{3/4}\, T^{3/2} \exp\left(-E_g / 2 k_B T\right)$ \cite{ashcroft}: when the PBEsol gap is comparable to $k_B T$, thermally excited intrinsic carriers vastly outnumber the nominal doping of $10^{18}~\text{carriers/cm}^{-3}$, both band edges contribute to the conductivity, and the extracted mass loses its meaning. Opening the gap with HSE06 suppresses $n_i$ exponentially and restores a well-defined, single-band-edge conductivity effective mass. For the 31 retained candidates with PBEsol gaps below \qty{0.3}{eV}, the mean gap opens from 0.13 to \qty{0.86}{eV}, and the largest effective-mass eigenvalue more than doubles (median factor of 2.7), in some cases growing by an order of magnitude: for example, in \ce{YN} the gap opens from 0.08 to \qty{1.06}{eV} while the electron mass increases from 0.03 to 0.26,$m_e$. For moderate-gap validation compounds the HSE06 gaps agree well with experiment (e.g. \ce{CdTe}: 1.54 vs \qty{1.5}{eV}; \ce{InP}: 1.43 vs \qty{1.35}{eV}), and the effective masses increase only moderately (median factor of 1.4), consistent with the near-proportionality of band-edge mass and gap expected from $k \cdot p$ arguments. 
The full HSE06-PBEsol parity plot for electrons is shown in Fig.~\ref{fig:electron_parity}, and the hole counterpart is provided in the Supplementary Material.

Re-ranking the candidates using the HSE06 effective masses confirms the necessity of higher-level functional corrections for final quantitative values, as well as the need to validate more than just a few top-ranked entries. The PBEsol and HSE06 orderings agree only moderately for holes (Spearman rank correlation coefficient $\rho = 0.79$) and poorly for electrons ($\rho = 0.18$), demonstrating that PBEsol ranking alone is not a reliable predictor of the hybrid result. This disagreement is concentrated among the lightest candidates: of the ten lightest materials at the PBEsol level, only four remain in the top ten under HSE06. Notably, for both electrons and holes, these four consist of the same narrow-gap group-IV/IV-VI chalcogenides (\ce{SnS}, \ce{SnTe}, \ce{GeSe}, and \ce{GeTe}), whose masses remain at or below $\approx 0.05\,m_e$ (with HSE06-to-PBEsol electron mass ratios of 1.8--3.8). These compounds therefore remain genuinely promising candidates. Conversely, the six materials that dropped off the shortlist (\ce{B6Si}, \ce{Li2MgSi}, \ce{Fe2GeTi}, \ce{GeMg2}, \ce{YN}, and \ce{As4Ge2Zn2}) all possess near-metallic PBEsol band gaps of \qtyrange{0.06}{0.15}{eV}. Once HSE06 opens these gaps by \qtyrange{0.4}{1.7}{eV}, their heaviest-direction electron mass increases by factors of $\approx 3\text{--}48$ (by nine-fold or more for five of the six). This confirms that their apparent low mass was an artifact of PBEsol band-gap underestimation rather than an intrinsic electronic property.

Taken together, these results illustrate how high-throughput Wannier-based workflows can uncover chemically diverse materials with ultralow, ultrahigh, or strongly anisotropic effective masses. The integration of dense Brillouin-zone sampling, automated transport analysis, and careful experimental literature validation provides a systematic route to discovering electronic materials beyond conventional chemistries. The candidate lists and prototype families identified here offer a starting point for targeted experimental synthesis and for higher-level theoretical studies of mobility, transport, and flat-band physics.

\section*{Methods}


While initial filtering relied on metadata from source databases to remove theoretical, high-pressure, and high-temperature entries, inconsistent labelling forced a rigorous manual verification of the primary experimental literature for all 520 final candidates. To ensure the reliability and practical applicability of the screened materials, we enforced following exclusion criteria. Structures were discarded if the original experimental reports indicated synthesis under extreme conditions (high pressure or non-standard temperatures) or if the materials were molecular crystals, organic frameworks, or contained disordered hydrogen.
Furthermore, we filtered for stability and handling safety; materials reported as hygroscopic (moisture-sensitive), photosensitive, explosive, or containing toxic cyanide groups were removed. Finally, as a quality control metric for the structural relaxation, we prioritized candidates where the volume difference between the DFT-relaxed and experimental structures was less than 5\%. 

To systematize the analysis of the screened candidates, we applied a heuristic grouping algorithm based on isovalent substitution. This approach reduces the high-dimensional compositional space into a set of chemical prototypes, allowing for the identification of trends across chemically distinct but electronically similar families. The clustering routine operates on the reduced chemical formula of each candidate. Each element is mapped to the representative element of its corresponding group in the periodic table (specifically, the period 2 element for $s$- and $p$-blocks, and the period 4 element for the $d$-block). For example, the chalcopyrite \ce{InAgS2} maps to the prototype \ce{CuBO2} (Group 13-11-16), while the Zintl phase \ce{CaMg2Bi2} maps to \ce{Be3N2} (Group 2-2-15).
Following this algorithmic reduction, prototypes were manually curated to align with established crystallographic conventions, and the corresponding conventional name is used in Table~\ref{tab:combined_materials} wherever one exists. For instance, the algorithmic prototype \ce{H2ZnF4} matches the stoichiometry and valence constraints of the Ruddlesden-Popper phase ($A_2BX_4$), and \ce{TiFN} represents the $\beta$-\ce{ZrNCl} layered nitride-halide ($MNX$) family. This classification scheme naturally separates isotropic 3D networks (e.g., III-V zincblendes) from low-dimensional systems (e.g., IV-VI bilayers), facilitating the separate analysis of anisotropic effective masses presented in Table~\ref{tab:combined_materials}.

\begin{acknowledgments}
We acknowledge financial support from the NCCR MARVEL (a National Centre of
  Competence in Research, funded by the Swiss National Science Foundation, grant
  No.
205602).
\end{acknowledgments}

\appendix

\bibliography{main_final_update}

\clearpage


\section*{Supplementary material}

# 1 Table of candidate materials

| Index | Formula | mc3d ID | Good material | Comment |
|-------|---------|---------|---------------|---------|
| 1 | **Cu2GeSe3** | mc3d-62111/pbesol-v1 | Yes | Used in solar cells |
| 2 | CdI2 | mc3d-13349/pbesol-v1 | No | Removed due to bad quality calculation |
| 3 | OsC2 | mc3d-49185/pbesol-v1 | No | First-principles calculations |
| 4 | **SiRh** | mc3d-13526/pbesol-v1 | Yes | Tested for superconductivity (susceptibility measurements) |
| 5 | **InAgS2** | mc3d-41262/pbesol-v1 | Yes | Synthetised at high temperature and pressure |
| 6 | BC2N | mc3d-67873/pbesol-v1 | No | Theoretical structure |
| 7 | AuBr | mc3d-29808/pbesol-v1 | No | On the edge of volume change criteria, stable between 353 and 470 K |
| 8 | **SnS** | mc3d-74977/pbesol-v1 | Yes | Thin layers |
| 9 | **Rb2Hg3Se4** | mc3d-4116/pbesol-v1 | Yes | |
| 10 | **K2(NbCl3)3** | mc3d-24544/pbesol-v1 | Yes | If the substance is left in the open air for a few days, slightly more dissolves |
| 11 | K2S5 | mc3d-31737/pbesol-v1 | No | Decomposes in air |
| 12 | LiGeTe2 | mc3d-64566/pbesol-v1 | No | Hygroscopic, reacts easily |
| 13 | **GeSe** | mc3d-28887/pbesol-v1 | Yes | Seems fine, but old USSR experimental paper which cant be found |
| 14 | **FeSi** | mc3d-12152/pbesol-v1 | Yes | This is the B20-type iron monosilicide ( 5-FeSi) with space group P213, it is stable under ambient conditions |
| 15 | **AgO** | mc3d-55179/pbesol-v1 | Yes | |
| 16 | **SnTe** | mc3d-55818/pbesol-v1 | Yes | Two phases described in the experimental paper |
| 17 | OsC | mc3d-4027/pbesol-v1 | No | First-principles calculations |
| 18 | **Sb2Rh** | mc3d-4805/pbesol-v1 | Yes | Seems fine but no experimental paper |
| 19 | **Ge(Te2As)2** | mc3d-53873/pbesol-v1 | Yes | Thermodynamically stable, hard to separate products |
| 20 | CdS | mc3d-46039/pbesol-v1 | No | Volume change too big + high pressure |
| 21 | **Te3As2** | mc3d-64991/pbesol-v1 | Yes | Metastable phase, may be maintained between 205 and 480 K, seems fine |
| 22 | CdS | mc3d-44515/pbesol-v1 | No | Volume change too big + high pressure |
| 23 | Te3As2 | mc3d-33661/pbesol-v1 | No | Stable below 205 K |
| 24 | CdAs | mc3d-11422/pbesol-v1 | No | High pressure |
| 25 | **SiB6** | mc3d-70485/pbesol-v1 | Yes | Seems fine, but old USSR experimental paper which cant be found |
| 26 | LiGaSi | mc3d-66491/pbesol-v1 | No | Volume change too big |
| 27 | **Ba2LiSi3** | mc3d-49519/pbesol-v1 | Yes | |
| 28 | GeRu | mc3d-65907/pbesol-v1 | Yes | Phase stable below 1670 K, then phase transition |
| 29 | **CaAgAs** | mc3d-75614/pbesol-v1 | Yes | Decomposed by dilute mineral acids |
| 30 | **Ba3SiI2** | mc3d-26769/pbesol-v1 | Yes | |
| 31 | RbCO | mc3d-3557/pbesol-v1 | No | Bad Wannier bands at CBM |
| 32 | FeS | mc3d-27727/pbesol-v1 | No | High pressure |
| 33 | PbO2 | mc3d-30209/pbesol-v1 | No | Volume change too big + high pressure |
| 34 | **Ba3GeI2** | mc3d-37814/pbesol-v1 | Yes | |
| 35 | **AgO** | mc3d-41965/pbesol-v1 | Yes | |
| 36 | **Li2MgSi** | mc3d-54484/pbesol-v1 | Yes | |
| 37 | **CaAgP** | mc3d-59762/pbesol-v1 | Yes | Soluble in dilute mineral acids |
| 38 | Ga3Ir | mc3d-5831/pbesol-v1 | Yes | Might be high-temperature phase acording to MPDS |
| 39 | **ZnSb** | mc3d-40969/pbesol-v1 | Yes | |
| 40 | **TiFe2Ge** | mc3d-65045/pbesol-v1 | Yes | |
| 41 | **In5CuS8** | mc3d-63221/pbesol-v1 | Yes | |
| 42 | Ta2Tl4S11 | mc3d-51930/pbesol-v1 | No | Removed due to bad quality calculation |
| 43 | **SiSbPt** | mc3d-8005/pbesol-v1 | Yes | |
| 44 | **Sc7CoCl12** | mc3d-48101/pbesol-v1 | Yes | Many similar structures in the experimental paper |

| Index | Formula | mc3d ID | Good material | Comment |
|---|---|---|---|---|
| 45 | **Mg2Ge** | mc3d-31279/pbesol-v1 | Yes | Many similar structures in the experimental paper |
| 46 | **B10(Pb2O7)3** | mc3d-31022/pbesol-v1 | Yes | |
| 47 | KCO | mc3d-16067/pbesol-v1 | No | Bad Wannier bands at CBM |
| 48 | **GeTe** | mc3d-41229/pbesol-v1 | Yes | Cant find the original experimental paper, might be a high-temperature phase |
| 49 | **ZrSe2** | mc3d-28539/pbesol-v1 | Yes | Many similar structures in the experimental paper |
| 50 | **YN** | mc3d-23080/pbesol-v1 | Yes | |
| 51 | Si3Mo | mc3d-4084/pbesol-v1 | No | High-temperature phase |
| 52 | **GeTe7As4** | mc3d-19550/pbesol-v1 | Yes | |
| 53 | CoAuO2 | mc3d-65474/pbesol-v1 | No | First-principles calculations |
| 54 | **BaSiC** | mc3d-8174/pbesol-v1 | Yes | |
| 55 | **Be5Pt** | mc3d-55967/pbesol-v1 | Yes | |
| 56 | **YAsSe** | mc3d-34738/pbesol-v1 | Yes | |
| 57 | BaMgSi | mc3d-21763/pbesol-v1 | No | First-principles calculations |
| 58 | SrZnF2 | mc3d-48231/pbesol-v1 | No | Volume change too big |
| 59 | **Ag3AuSe2** | mc3d-1202/pbesol-v1 | Yes | |
| 60 | **ReTeS** | mc3d-53704/pbesol-v1 | Yes | |
| 61 | **Ba3Nb2O8** | mc3d-19377/pbesol-v1 | Yes | |
| 62 | **Na(LiSi2)3** | mc3d-13089/pbesol-v1 | Yes | |
| 63 | **Ba2NF** | mc3d-5654/pbesol-v1 | Yes | |
| 64 | Be3P2 | mc3d-74003/pbesol-v1 | No | First-principles calculations |
| 65 | **K2Ge2Se5** | mc3d-35753/pbesol-v1 | Yes | |
| 66 | **GaAs** | mc3d-13434/pbesol-v1 | Yes | |
| 67 | **ZnGeAs2** | mc3d-7337/pbesol-v1 | Yes | |
| 68 | **FeTe2** | mc3d-9510/pbesol-v1 | Yes | |
| 69 | Cd2SnO4 | mc3d-10402/pbesol-v1 | No | Volume change too big |
| 70 | **Cd(InSe2)2** | mc3d-74808/pbesol-v1 | Yes | |
| 71 | GaAs | mc3d-37196/pbesol-v1 | No | Theoretical structure |
| 72 | SnO2 | mc3d-72875/pbesol-v1 | No | First-principles calculation |
| 73 | Na3As | mc3d-71481/pbesol-v1 | No | Large change in volume, high pressure phase |
| 74 | LiCdSb | mc3d-29506/pbesol-v1 | No | Theoretical structure |
| 75 | **TaTlSe3** | mc3d-12745/pbesol-v1 | Yes | |
| 76 | BaPbO3 | mc3d-44950/pbesol-v1 | No | High-temperature phase at 573 K |
| 77 | **Te8RuCl2** | mc3d-33154/pbesol-v1 | Yes | |
| 78 | **Ge2Te5As2** | mc3d-8457/pbesol-v1 | Yes | |
| 79 | **InAgTe2** | mc3d-26941/pbesol-v1 | Yes | |
| 80 | N12Ne7 | mc3d-54807/pbesol-v1 | No | Large change in volume |
| 81 | BaCl2 | mc3d-23592/pbesol-v1 | No | High temperature phase |
| 82 | **LiMgBi** | mc3d-62475/pbesol-v1 | Yes | |
| 83 | CaSeO2 | mc3d-75012/pbesol-v1 | No | Volume change too big |
| 84 | Al2HgSe4 | mc3d-24891/pbesol-v1 | No | Theoretical structure |
| 85 | **AgBiSe2** | mc3d-40857/pbesol-v1 | Yes | |
| 86 | **GaCuSe2** | mc3d-16413/pbesol-v1 | Yes | |
| 87 | **Sr(MgBi)2** | mc3d-19595/pbesol-v1 | Yes | |
| 88 | GeS | mc3d-70347/pbesol-v1 | No | Volume change too big + high pressure |
| 89 | **GaCuTe2** | mc3d-26212/pbesol-v1 | Yes | |
| 90 | **CdSe** | mc3d-58436/pbesol-v1 | Yes | |
| 91 | RhN3 | mc3d-24235/pbesol-v1 | No | First-principles calculations |
| 92 | **SiRu** | mc3d-61851/pbesol-v1 | Yes | |
| 93 | **CdSnP2** | mc3d-13158/pbesol-v1 | Yes | |
| 94 | MnAlGe | mc3d-51680/pbesol-v1 | No | Volume change too big |
| 95 | **Mg2Si** | mc3d-35349/pbesol-v1 | Yes | |
| 96 | **CsCu5Se3** | mc3d-69863/pbesol-v1 | Yes | |
| 97 | **InP** | mc3d-72275/pbesol-v1 | Yes | |
| 98 | **Si(Hg2Se3)2** | mc3d-52332/pbesol-v1 | Yes | |
| 99 | HfC | mc3d-38229/pbesol-v1 | No | Volume change too big |
| 100 | **CdSiAs2** | mc3d-67280/pbesol-v1 | Yes | |
| 101 | **Li2MgSi** | mc3d-29074/pbesol-v1 | Yes | |
| 102 | **CdTe** | mc3d-64416/pbesol-v1 | Yes | |

| Index | Formula | mc3d ID | Good material | Comment |
|-------|---------|---------|---------------|---------|
| 103 | Na3Sb | mc3d-76469/pbesol-v1 | No | High-pressure phase stable above 2.3 GPa |
| 104 | **CdTe** | mc3d-62382/pbesol-v1 | Yes | |
| 105 | **Ca(GaP)2** | mc3d-20007/pbesol-v1 | Yes | |
| 106 | InP | mc3d-62948/pbesol-v1 | No | Theoretical structure |
| 107 | ZnCd3Se4 | mc3d-69558/pbesol-v1 | No | Theoretical structure |
| 108 | **Mn2MoP12** | mc3d-26270/pbesol-v1 | Yes | |
| 109 | Bi2Se3 | mc3d-39833/pbesol-v1 | No | High volume change percentage |
| 110 | **CdSe** | mc3d-31089/pbesol-v1 | Yes | |
| 111 | **AgBiSe2** | mc3d-36262/pbesol-v1 | Yes | |
| 112 | RbCuTe | mc3d-61752/pbesol-v1 | No | High pressure structure |
| 113 | Na3As | mc3d-12466/pbesol-v1 | No | High volume change percentage |
| 114 | **GaAgSe2** | mc3d-40127/pbesol-v1 | Yes | |
| 115 | **Li2AgSb** | mc3d-50673/pbesol-v1 | Yes | |
| 116 | **LiZnAs** | mc3d-36893/pbesol-v1 | Yes | |
| 117 | **In2HgS4** | mc3d-61190/pbesol-v1 | Yes | |
| 118 | ScSnAu | nan | No | MPDS entry missing |
| 119 | **Sn(HgSe2)2** | mc3d-16398/pbesol-v1 | Yes | Seems fine, but no experimental paper |
| 120 | Na3InAs2 | mc3d-26917/pbesol-v1 | Yes | |
| 121 | CdTe | mc3d-19583/pbesol-v1 | Yes | |
| 122 | **Hg2GeSe4** | mc3d-21668/pbesol-v1 | Yes | |
| 123 | VCu3O4 | mc3d-45924/pbesol-v1 | Yes | |
| 124 | **Sr2Zn2As3** | mc3d-60357/pbesol-v1 | Yes | |
| 125 | GaTe | mc3d-36254/pbesol-v1 | Yes | Old USSR experimental paper, but seems fine |
| 126 | **In2S3** | mc3d-79779/pbesol-v1 | Yes | |
| 127 | ZnSnP2 | mc3d-22927/pbesol-v1 | No | High temperature structure |
| 128 | **Al2Ru** | mc3d-40278/pbesol-v1 | Yes | |
| 129 | **Al4SiC4** | mc3d-57354/pbesol-v1 | Yes | |
| 130 | **PdSe** | mc3d-27207/pbesol-v1 | Yes | |
| 131 | FeTeAs | mc3d-11361/pbesol-v1 | Yes | |
| 132 | Fe2Ge3 | mc3d-43320/pbesol-v1 | Yes | |
| 133 | **Ba(MgBi)2** | mc3d-31118/pbesol-v1 | Yes | |
| 134 | ZnTe | mc3d-68088/pbesol-v1 | No | Theoretical structure |
| 135 | **Bi2Te2S** | mc3d-56846/pbesol-v1 | Yes | |
| 136 | Cs2Te5 | mc3d-45308/pbesol-v1 | No | Decomposes easily |
| 137 | Ca6Ge2O | mc3d-48159/pbesol-v1 | No | DFT data |
| 138 | **Mg3Sb2** | mc3d-36823/pbesol-v1 | Yes | |
| 139 | Zn3CdSe4 | mc3d-32155/pbesol-v1 | No | Theoretical structure |
| 140 | CdI2 | mc3d-5628/pbesol-v1 | Yes | We checked other nH polytypes earlier |
| 141 | **GaReSi** | mc3d-14329/pbesol-v1 | Yes | |
| 142 | Nb2GeTe4 | mc3d-5989/pbesol-v1 | Yes | Seems fine, but no experimental paper |
| 143 | **MgGeAs2** | mc3d-9557/pbesol-v1 | Yes | |
| 144 | ZnSe | mc3d-56292/pbesol-v1 | No | High pressure structure |
| 145 | **Ti2O3** | mc3d-23607/pbesol-v1 | Yes | High pressure experimental paper, but includes the low pressure structure as well. I think it is fine, but MPDS says: high-pressure phase stable above 20 GPa |
| 146 | **Mg3As2** | mc3d-3745/pbesol-v1 | Yes | |
| 147 | **Y2ZnSe4** | nan | Yes | No experimental paper available, old USSR work |
| 148 | WCl4 | mc3d-14132/pbesol-v1 | No | Sensitive to humidity according to MPDS |
| 149 | Al2CdSe4 | mc3d-7819/pbesol-v1 | No | Theoretical structure |
| 150 | SnC | mc3d-73725/pbesol-v1 | No | Theoretical structure |
| 151 | ZnCdSe2 | mc3d-5145/pbesol-v1 | No | Theoretical structure |
| 152 | **Ge4Te7As2** | mc3d-19656/pbesol-v1 | Yes | |
| 153 | Sr5(SnAs3)2 | mc3d-49495/pbesol-v1 | No | Compound Eu5Sn2As6 showed good stability in air; however, the crystals of Sr5Sn2As6 decomposed quickly when exposed to air or moisture |

| Index | Formula | mc3d ID | Good material | Comment |
|---|---|---|---|---|
| 154 | **YSbPt** | mc3d-64975/pbesol-v1 | Yes | |
| 155 | **ZnTe** | mc3d-75435/pbesol-v1 | Yes | |
| 156 | **Na3P** | mc3d-60261/pbesol-v1 | Yes | |
| 157 | **CdGeP2** | mc3d-62820/pbesol-v1 | Yes | |
| 158 | **LiCdP** | mc3d-11960/pbesol-v1 | Yes | |
| 159 | **NaZnAs** | mc3d-53666/pbesol-v1 | Yes | |
| 160 | **Al5C3N** | mc3d-73948/pbesol-v1 | Yes | |
| 161 | **InAgS2** | mc3d-54322/pbesol-v1 | Yes | |
| 162 | **Sn3N4** | mc3d-78193/pbesol-v1 | Yes | |
| 163 | **Na3Sb** | mc3d-63487/pbesol-v1 | Yes | |
| 164 | ZnTe | mc3d-56020/pbesol-v1 | No | High pressure structure |
| 165 | **ZnTe** | mc3d-22365/pbesol-v1 | Yes | |
| 166 | **Ag3PO4** | mc3d-9267/pbesol-v1 | Yes | No experimental paper, but seems fine |
| 167 | Se | mc3d-66291/pbesol-v1 | No | 11% Change in volume |
| 168 | **Y6OsI10** | mc3d-70477/pbesol-v1 | Yes | |
| 169 | **Ca5(GaAs3)2** | mc3d-58126/pbesol-v1 | Yes | |
| 170 | **BaNaBi** | mc3d-17698/pbesol-v1 | Yes | |
| 171 | **Ga5CuTe8** | mc3d-4160/pbesol-v1 | Yes | |
| 172 | **In6S7** | mc3d-31006/pbesol-v1 | Yes | Solar energy applications |
| 173 | **Na5SnAs3** | mc3d-68697/pbesol-v1 | Yes | |
| 174 | FeO2 | mc3d-31148/pbesol-v1 | No | High volume change percentage |
| 175 | **Rb2PtI6** | mc3d-6673/pbesol-v1 | Yes | |
| 176 | **Ca(MgBi)2** | mc3d-14953/pbesol-v1 | Yes | |
| 177 | **RbSrCl3** | mc3d-15465/pbesol-v1 | Yes | |
| 178 | Be(BC)2 | mc3d-27846/pbesol-v1 | No | Beryllium and beryllium compounds are highly toxic and may act as human carcinogens |
| 179 | **AgSbO3** | mc3d-50569/pbesol-v1 | Yes | |
| 180 | NbO2 | mc3d-62310/pbesol-v1 | No | High pressure strucutre |
| 181 | **InGaSe3** | mc3d-46231/pbesol-v1 | Yes | |
| 182 | CdI2 | mc3d-21208/pbesol-v1 | Yes | Long unit cell |
| 183 | **CdS** | mc3d-51556/pbesol-v1 | Yes | |
| 184 | **In5AgSe8** | mc3d-55445/pbesol-v1 | Yes | |
| 185 | Ca2As3 | mc3d-65887/pbesol-v1 | No | Very sensitive to moisture and decomposes when exposed to open air |
| 186 | **Zn3(InS3)2** | mc3d-34052/pbesol-v1 | Yes | Seems fine, but no experimental experimental paper, old USSR work |
| 187 | **ZnSe** | mc3d-11086/pbesol-v1 | Yes | |
| 188 | **KNa2Sb** | mc3d-18773/pbesol-v1 | Yes | |
| 189 | **Zn(InS2)2** | mc3d-70423/pbesol-v1 | Yes | |
| 190 | **As2Pt** | mc3d-16865/pbesol-v1 | Yes | |
| 191 | **BaSi9Rh2** | mc3d-31647/pbesol-v1 | Yes | Nice thermoelectric material, but all the measurements were done in argon |
| 192 | **Y6RuI10** | mc3d-44347/pbesol-v1 | Yes | |
| 193 | **In5CuSe8** | mc3d-17204/pbesol-v1 | Yes | |
| 194 | **Ge** | mc3d-1639/pbesol-v1 | Yes | |
| 195 | **Cd(InTe2)2** | mc3d-34912/pbesol-v1 | Yes | |
| 196 | **AlCuSe2** | mc3d-5930/pbesol-v1 | Yes | |
| 197 | **Cd(InS2)2** | mc3d-65185/pbesol-v1 | Yes | |
| 198 | PbSe | mc3d-66459/pbesol-v1 | No | High volume change percentage |
| 199 | **Cd4SiSe6** | mc3d-65435/pbesol-v1 | Yes | |
| 200 | **BaSbTe3** | mc3d-29622/pbesol-v1 | Yes | |
| 201 | **Sr(As3Pt2)2** | mc3d-57759/pbesol-v1 | Yes | |
| 202 | KMgF3 | mc3d-14794/pbesol-v1 | No | High volume change percentage |
| 203 | **AlAgTe2** | nan | Yes | |
| 204 | **AlCuTe2** | mc3d-64297/pbesol-v1 | Yes | |
| 205 | KTl | mc3d-73577/pbesol-v1 | No | Sensitive to air and moisture |
| 206 | **TlInS2** | mc3d-18900/pbesol-v1 | Yes | Seems fine, but no experimental paper, old USSR work |
| 207 | **GaSe** | mc3d-50014/pbesol-v1 | Yes | |
| 208 | **CuBr** | mc3d-866/pbesol-v1 | Yes | |
| 209 | **CdS** | mc3d-23672/pbesol-v1 | Yes | |

4

| Index | Formula | mc3d ID | Good material | Comment |
|-------|---------|---------|---------------|---------|
| 210 | **GaSe** | mc3d-47142/pbesol-v1 | Yes | |
| 211 | **InSe** | mc3d-4386/pbesol-v1 | Yes | |
| 212 | **In2HgTe4** | mc3d-5348/pbesol-v1 | Yes | |
| 213 | **NaCuTe** | mc3d-76518/pbesol-v1 | Yes | |
| 214 | **Cd(InSe2)2** | mc3d-52014/pbesol-v1 | Yes | |
| 215 | ZnO | mc3d-3672/pbesol-v1 | No | Large change in volume |
| 216 | **Cd(InSe2)2** | mc3d-60099/pbesol-v1 | Yes | |
| 217 | **GaSe** | mc3d-60842/pbesol-v1 | Yes | |
| 218 | **K2NaAs** | mc3d-45533/pbesol-v1 | Yes | |
| 219 | **Zn(InS2)2** | mc3d-40561/pbesol-v1 | Yes | |
| 220 | **NbS3** | mc3d-971/pbesol-v1 | Yes | Old USSR experimental paper |
| 221 | **GaSe** | mc3d-58792/pbesol-v1 | Yes | |
| 222 | **TlInTe2** | mc3d-54389/pbesol-v1 | Yes | Seems fine, but no experimental paper |
| 223 | **ScNiBi** | mc3d-10910/pbesol-v1 | Yes | Half Heusler. No experimental paper, but seems fine |
| 224 | Mg2C | mc3d-26943/pbesol-v1 | No | High pressure structure |
| 225 | **PbS** | mc3d-76716/pbesol-v1 | Yes | No experimental paper, called high-pressure, but according to mpds it is stable below 2.43GPa, so at ambient pressure it should be stable |
| 226 | **LiZnN** | mc3d-23550/pbesol-v1 | Yes | |
| 227 | **InSe** | mc3d-49189/pbesol-v1 | Yes | |
| 228 | SrAl6O9 | mc3d-71411/pbesol-v1 | No | MPDS formula is Sr0.89Eu0.11Al6O10, however structure is missing Eu. Experimental paper is about Europium (Eu), but no Eu in the structure |
| 229 | **In2HgSe4** | mc3d-73776/pbesol-v1 | Yes | |
| 230 | SiSn | mc3d-19663/pbesol-v1 | No | Theoretical structure |
| 231 | ZnCd3S4 | mc3d-10905/pbesol-v1 | No | Theoretical structure |
| 232 | CoSbS | nan | No | ICSD structure deleted. No results found for the given search criteria! |
| 233 | ZnO | mc3d-38845/pbesol-v1 | No | Theoretical structure |
| 234 | **Cd(InSe2)2** | mc3d-6700/pbesol-v1 | Yes | |
| 235 | **Cs3Bi** | mc3d-63109/pbesol-v1 | Yes | |
| 236 | **GaGeTe** | mc3d-14913/pbesol-v1 | Yes | |
| 237 | **TePb** | mc3d-15774/pbesol-v1 | Yes | |
| 238 | SnO2 | mc3d-68897/pbesol-v1 | No | High pressure structure |
| 239 | RbGeI3 | mc3d-59053/pbesol-v1 | No | Volume change too big |
| 240 | **NaSrAs** | mc3d-51397/pbesol-v1 | Yes | |
| 241 | **KCdAs** | mc3d-73007/pbesol-v1 | Yes | |
| 242 | **SiPAu** | mc3d-7637/pbesol-v1 | Yes | |
| 243 | SrGeO3 | mc3d-74559/pbesol-v1 | No | High-temperature high-pressure phase stable above 5.0 GPa at 983-1513 K |
| 244 | **Al2HgS4** | mc3d-7453/pbesol-v1 | Yes | |
| 245 | PbS | mc3d-7518/pbesol-v1 | No | High volume change percentage |
| 246 | **CsK2Sb** | mc3d-29433/pbesol-v1 | Yes | |
| 247 | **BaSnO3** | mc3d-61206/pbesol-v1 | Yes | |
| 248 | **In2Se3** | mc3d-3253/pbesol-v1 | Yes | |
| 249 | PbS | mc3d-24444/pbesol-v1 | No | High volume change percentage |
| 250 | ZnO | mc3d-69913/pbesol-v1 | No | Theoretical structure |
| 251 | **Cd(InSe2)2** | mc3d-37892/pbesol-v1 | Yes | |
| 252 | **Sb2TeSe2** | mc3d-6753/pbesol-v1 | Yes | Seems fine, but no experimental paper |
| 253 | **GaCuS2** | mc3d-54526/pbesol-v1 | Yes | |
| 254 | **Cd(InTe2)2** | mc3d-57872/pbesol-v1 | Yes | |
| 255 | **AlAgSe2** | mc3d-23865/pbesol-v1 | Yes | |
| 256 | TePb | mc3d-64878/pbesol-v1 | No | High pressure structure |
| 257 | **Zn(InSe2)2** | mc3d-6119/pbesol-v1 | Yes | |
| 258 | ZnO | mc3d-44016/pbesol-v1 | No | Theoretical structure |
| 259 | ZnSnS3 | mc3d-2056/pbesol-v1 | No | Theoretical structure |
| 260 | **GaAgS2** | mc3d-41467/pbesol-v1 | Yes | |
| 261 | **Y(ZnP)3** | mc3d-3229/pbesol-v1 | Yes | |
| 262 | **Sr2CdAs2** | mc3d-3845/pbesol-v1 | Yes | |

| Index | Formula | mc3d ID | Good material | Comment |
|-------|---------|---------|---------------|---------|
| 263 | Zn3CdS4 | mc3d-6814/pbesol-v1 | No | Theoretical structure |
| 264 | **Cd4GeSe6** | mc3d-61300/pbesol-v1 | Yes | |
| 265 | **Na5SiAs3** | mc3d-23836/pbesol-v1 | Yes | |
| 266 | **NbCuN2** | mc3d-55669/pbesol-v1 | Yes | |
| 267 | SnO2 | nan | No | 12% Change in volume, original MPDS entry missing |
| 268 | **K3In2As3** | mc3d-50194/pbesol-v1 | Yes | |
| 269 | AuCN | mc3d-61965/pbesol-v1 | No | Cyanide, toxic |
| 270 | KGeI3 | mc3d-44962/pbesol-v1 | No | First-principles calculations |
| 271 | **Cd3TeO6** | mc3d-52139/pbesol-v1 | Yes | |
| 272 | PbS | mc3d-62703/pbesol-v1 | No | High volume change percentage |
| 273 | SnO2 | mc3d-30492/pbesol-v1 | No | High pressure structure |
| 274 | **K3P** | mc3d-47682/pbesol-v1 | Yes | |
| 275 | ZnCdS2 | mc3d-74103/pbesol-v1 | No | Theoretical structure |
| 276 | **Al2ZnSe4** | mc3d-47431/pbesol-v1 | Yes | |
| 277 | **K3Sb** | mc3d-40580/pbesol-v1 | Yes | Thin film only confirmed, photoemiter |
| 278 | **Zn(InTe2)2** | mc3d-59767/pbesol-v1 | Yes | |
| 279 | **Sb3Ir** | mc3d-34197/pbesol-v1 | Yes | |
| 280 | CuBr | mc3d-5894/pbesol-v1 | No | Large change in volume, high-temperature phase stable between 658 and 743 K |
| 281 | **Sb2Te4Pb** | mc3d-58478/pbesol-v1 | Yes | |
| 282 | SiO2 | mc3d-5355/pbesol-v1 | No | |
| 283 | **Na2LiN** | mc3d-15837/pbesol-v1 | Yes | |
| 284 | **CdSnO3** | mc3d-79129/pbesol-v1 | Yes | No experimental paper at mcsd, but checkout ICSD 180402 and 165167 looks good |
| 285 | ZnO | mc3d-56509/pbesol-v1 | No | Theoretical structure |
| 286 | GeO2 | mc3d-4424/pbesol-v1 | No | Large change in volume |
| 287 | **Ga2Se3** | mc3d-58700/pbesol-v1 | Yes | |
| 288 | BiB | mc3d-24918/pbesol-v1 | No | Theoretical structure |
| 289 | In2O3 | mc3d-51848/pbesol-v1 | No | High pressure |
| 290 | RbW3Br7 | mc3d-46975/pbesol-v1 | No | Removed due to bad quality calculation |
| 291 | **Al2HgTe4** | mc3d-19061/pbesol-v1 | Yes | |
| 292 | **K4ZnBi2** | mc3d-42391/pbesol-v1 | Yes | |
| 293 | **MgTe** | mc3d-7683/pbesol-v1 | Yes | |
| 294 | **ZrGeTe4** | mc3d-49004/pbesol-v1 | Yes | |
| 295 | SnO2 | mc3d-27863/pbesol-v1 | No | High pressure and temperature |
| 296 | **ZnO** | mc3d-31594/pbesol-v1 | Yes | |
| 297 | TePb | mc3d-49931/pbesol-v1 | No | High pressure structure |
| 298 | **ZnS** | mc3d-72223/pbesol-v1 | Yes | |
| 299 | **ZnS** | mc3d-38390/pbesol-v1 | Yes | Nanoparticles, optoelectronic properties |
| 300 | **ZnS** | mc3d-66021/pbesol-v1 | Yes | |
| 301 | AgI | mc3d-37668/pbesol-v1 | No | Might decompose under light |
| 302 | **Hg(SbO3)2** | mc3d-64934/pbesol-v1 | Yes | No experimental paper, but seems ok |
| 303 | PbSe | mc3d-36258/pbesol-v1 | No | High pressure |
| 304 | BaF2 | mc3d-58341/pbesol-v1 | No | High percentage change in volume |
| 305 | **ZnS** | mc3d-77731/pbesol-v1 | Yes | |
| 306 | CsPbI3 | mc3d-8755/pbesol-v1 | No | High-temperature phase stable above 563 K |
| 307 | **Zn(InSe2)2** | mc3d-56951/pbesol-v1 | Yes | |
| 308 | TiPO4 | mc3d-52975/pbesol-v1 | No | High volume change percentage |
| 309 | **TlBiS2** | mc3d-36026/pbesol-v1 | Yes | No experimental paper available, old USSR work, but seems fine |
| 310 | AgI | mc3d-56189/pbesol-v1 | No | It's a well known chemical in photography, so it's light sensitive |
| 311 | **Zn(GaSe2)2** | mc3d-44192/pbesol-v1 | Yes | |
| 312 | **Y2CdSe4** | mc3d-38949/pbesol-v1 | Yes | |
| 313 | **ZnS** | mc3d-15053/pbesol-v1 | Yes | |

| Index | Formula | mc3d ID | Good material | Comment |
|-------|---------|---------|---------------|---------|
| 314 | **CdSnO3** | mc3d-60866/pbesol-v1 | Yes | No experimental experimental paper, transforms to a perovskite phase at high temperature and high pressure, according to other experimental paper |
| 315 | **Ga2HgTe4** | mc3d-63007/pbesol-v1 | Yes | |
| 316 | **K5SnAs3** | mc3d-35830/pbesol-v1 | Yes | |
| 317 | KMgAs | mc3d-42168/pbesol-v1 | No | Sensitive to air |
| 318 | **Sc2CdSe4** | mc3d-57266/pbesol-v1 | Yes | |
| 319 | **ZnS** | mc3d-79026/pbesol-v1 | Yes | Very brief experimental paper |
| 320 | **GaN** | mc3d-3763/pbesol-v1 | Yes | |
| 321 | ScBiO3 | mc3d-77301/pbesol-v1 | No | DFT structure |
| 322 | **Mg3N2** | mc3d-52431/pbesol-v1 | Yes | |
| 323 | AgCl | mc3d-27640/pbesol-v1 | No | No experimental paper, but this material is light sensitive |
| 324 | AgI | mc3d-57698/pbesol-v1 | No | Light sensitive |
| 325 | **K3Ga3As4** | mc3d-38905/pbesol-v1 | Yes | |
| 326 | AgI | mc3d-77283/pbesol-v1 | No | Light sensitive |
| 327 | **ZnS** | mc3d-17299/pbesol-v1 | Yes | |
| 328 | **In2Ge2O7** | mc3d-26750/pbesol-v1 | Yes | |
| 329 | ZnO | nan | No | No entry at mpds |
| 330 | **ZnS** | mc3d-44841/pbesol-v1 | Yes | ZnS,there's lots of it |
| 331 | **Ge4Se9** | mc3d-2375/pbesol-v1 | Yes | |
| 332 | **ZnS** | mc3d-9909/pbesol-v1 | Yes | |
| 333 | **In6TeO12** | mc3d-10799/pbesol-v1 | Yes | |
| 334 | AgI | mc3d-23565/pbesol-v1 | No | Light sensitive |
| 335 | **Na5GeAs3** | mc3d-10548/pbesol-v1 | Yes | |
| 336 | **NaMgAs** | mc3d-27162/pbesol-v1 | Yes | |
| 337 | **K3InP2** | mc3d-42415/pbesol-v1 | Yes | |
| 338 | **Ca2CdAs2** | mc3d-62645/pbesol-v1 | Yes | |
| 339 | **As3Ir** | mc3d-26954/pbesol-v1 | Yes | |
| 340 | AgBr | mc3d-75102/pbesol-v1 | No | Might decompose under light |
| 341 | **Tl5Te2I** | mc3d-49465/pbesol-v1 | Yes | |
| 342 | MgTe | nan | No | No entry at mpds |
| 343 | **ZnS** | mc3d-26001/pbesol-v1 | Yes | ZnS, |
| 344 | **ZnS** | mc3d-30743/pbesol-v1 | Yes | |
| 345 | **ZnS** | mc3d-43282/pbesol-v1 | Yes | ZnS, |
| 346 | **Cd2Sb2O7** | mc3d-23813/pbesol-v1 | Yes | Perovskite |
| 347 | **Ga2HgSe4** | mc3d-35179/pbesol-v1 | Yes | |
| 348 | **ZnS** | mc3d-10851/pbesol-v1 | Yes | ZnS, |
| 349 | InSbO4 | mc3d-43369/pbesol-v1 | No | No experimental paper, according to mpds it's high pressure phase |
| 350 | NaAlSe2 | mc3d-62205/pbesol-v1 | No | Sensitive to air and humidity |
| 351 | **Ca2CdAs2** | mc3d-5545/pbesol-v1 | Yes | |
| 352 | B3C10N3 | mc3d-56069/pbesol-v1 | No | High volume change percentage |
| 353 | **Cd5(PBr2)2** | mc3d-19389/pbesol-v1 | Yes | Unavailable russian experimental paper |
| 354 | Mg3AsN | mc3d-26894/pbesol-v1 | No | Sensitive to air |
| 355 | ZnS | mc3d-538/pbesol-v1 | No | ZnS, 16H2 phase |
| 356 | Mg3N2 | mc3d-563/pbesol-v1 | No | High percentage change in volume |
| 357 | **Cs3Sb** | mc3d-27595/pbesol-v1 | Yes | |
| 358 | **GaN** | mc3d-6113/pbesol-v1 | Yes | |
| 359 | AgBr | mc3d-72945/pbesol-v1 | No | High percentage change in volume |
| 360 | **ZnS** | mc3d-19733/pbesol-v1 | Yes | |
| 361 | SrZrN2 | mc3d-2911/pbesol-v1 | No | Sensitive to air and humidity |
| 362 | MgSe | mc3d-6754/pbesol-v1 | No | Theoretical experimental paper |
| 363 | **ZnS** | mc3d-40532/pbesol-v1 | Yes | ZnS,, |
| 364 | **SnPtSe** | mc3d-57597/pbesol-v1 | Yes | |
| 365 | TlBr | mc3d-79451/pbesol-v1 | No | 5% Change in volume |
| 366 | K5CuSb2 | mc3d-72315/pbesol-v1 | No | Sensitive to humidity |
| 367 | Zn2SnO4 | mc3d-46906/pbesol-v1 | No | Theoretical work |
| 368 | **Na3AlAs2** | mc3d-28209/pbesol-v1 | Yes | |
| 369 | **K2Cd3Se4** | mc3d-31467/pbesol-v1 | Yes | |
| 370 | **ZnS** | mc3d-47544/pbesol-v1 | Yes | ZnS, |

| Index | Formula | mc3d ID | Good material | Comment |
|-------|---------|---------|---------------|---------|
| 371 | Ga2Ru | mc3d-37718/pbesol-v1 | No | DFT data |
| 372 | **Cu(PtS2)2** | mc3d-29365/pbesol-v1 | Yes | |
| 373 | **Cd(GaSe2)2** | mc3d-17472/pbesol-v1 | Yes | |
| 374 | **Hg(AsO3)2** | mc3d-47991/pbesol-v1 | Yes | |
| 375 | AlBiO3 | mc3d-51594/pbesol-v1 | No | First priciples study |
| 376 | ZnO | mc3d-77328/pbesol-v1 | No | First principles study |
| 377 | **ZnS** | mc3d-61234/pbesol-v1 | Yes | ZnS |
| 378 | CuP10 | mc3d-21689/pbesol-v1 | No | Removed due to bad quality calculation |
| 379 | CdGeO3 | mc3d-71681/pbesol-v1 | No | High pressure |
| 380 | **Rb2SnAs2** | mc3d-55124/pbesol-v1 | Yes | |
| 381 | **Ga2TeO6** | mc3d-47411/pbesol-v1 | Yes | Experimental paper about magnetic phase transitions |
| 382 | **ZnS** | mc3d-41344/pbesol-v1 | Yes | ZnS |
| 383 | AgBr | mc3d-38442/pbesol-v1 | No | Light sensitive, but less than AgI if I recall correctly, experimental paper focused on different materials |
| 384 | **Ca(CuS)2** | mc3d-54735/pbesol-v1 | Yes | |
| 385 | ZnSnO3 | nan | No | No entry at mpds |
| 386 | SnO2 | mc3d-21788/pbesol-v1 | No | High pressure |
| 387 | Sc4C3 | mc3d-19756/pbesol-v1 | No | High pressure |
| 388 | **Cd4GeS6** | mc3d-24877/pbesol-v1 | Yes | |
| 389 | **Sr3SbN** | mc3d-3209/pbesol-v1 | Yes | |
| 390 | **ScBiPd** | mc3d-62829/pbesol-v1 | Yes | Seems ok, but might be toxic |
| 391 | Si | mc3d-75862/pbesol-v1 | No | DFT data |
| 392 | SnO2 | mc3d-54082/pbesol-v1 | No | Topological experimental paper |
| 393 | **In2TeO6** | mc3d-72313/pbesol-v1 | Yes | No experimental paper available, seems ok |
| 394 | **Zn2BrN** | mc3d-78884/pbesol-v1 | Yes | |
| 395 | **CaSnO3** | mc3d-36414/pbesol-v1 | Yes | Very old experimental paper 1926, but seems ok, perovskite material |
| 396 | BaO2 | mc3d-41801/pbesol-v1 | No | Moisture sensitive |
| 397 | TeBr6 | mc3d-12993/pbesol-v1 | No | High volume change percentage |
| 398 | **Mg3SbN** | mc3d-10736/pbesol-v1 | Yes | |
| 399 | **CdIBr** | mc3d-15171/pbesol-v1 | Yes | Missing experimental paper but seems fine |
| 400 | CsAu | mc3d-35362/pbesol-v1 | No | High volume change percentage |

## 2 Additional information and datasheets

For each material identified as a good candidate in the table above, we provide a detailed datasheet summarizing its general and electronic properties. Each datasheet contains a table with general properties, where the hole/electron effective mass stability is defined as the relative difference between the effective mass calculated at doping levels of $10^{20}$ and $10^{18}$ cm$^{-3}$:

$$\frac{m^*_{e20} - m^*_{e18}}{\min(m^*_{e20}, m^*_{e18})}$$

This value is highlighted in red when it exceeds 50%. To the right of this table, we present the eigenvalues of the effective mass tensor at three different doping levels. The datasheet also includes plots showing the band structure, the effective mass plotted against doping concentration, and the Density of States (DOS), both part corresponding to the bandstructure and enlarged around the Valence Band Maximum (VBM) and Conduction Band Minimum (CBM). In the enlarged DOS plots, rectangles indicate the energy range of the Fermi-Dirac function where $10^{-3} < f(T = 300K) < 1 - 10^{-3}$. Missing rectangles in some plots are the result of smearing used in the calculations causing non-zero DOS values between VBM and CBM.

# Cu$_2$GeSe$_3$

| Formula IUPAC | Cu$_2$GeSe$_3$ | Num. atoms | 12 |
|---|---|---|---|
| Formula Hill | Cu$_2$Ge$_2$Se$_{90}$ | Space group | Cc (9) |
| Boltzmann UUID | e3d9693f7-900b-4baa-9a3f-27360ceb14f3 | Source database | MC3D |
| W90optwkc UUID | 75d9b477-b1eb-4f0-a021-b850403f28ea | MC3D ID | mp3d-62111/pbesol-v1 |
| Electron stability | **10000.00%** | Bandgap | 0.07 eV |
| Hole stability | **9690.00%** | Band distance | 9.9 meV |

**Eigenvalues of the effective mass tensors in electron mass units [$m_e$].**

Absolute value shown for three doping levels: $10^{18}$, $10^{19}$, $10^{20}$ for electrons and holes.

| Effective masses | Doping $10^{18}$ | Doping $10^{19}$ | Doping $10^{20}$ |
|---|---|---|---|
| electron | (0.00011, **0.00017**, 0.00012) | (0.00011, **0.00017**, 0.00012) | (0.011, **0.0017**, 0.012) |
| hole | (0.00011, **0.00017**, 0.00012) | (0.00011, **0.00017**, 0.00012) | (0.011, **0.0017**, 0.012) |

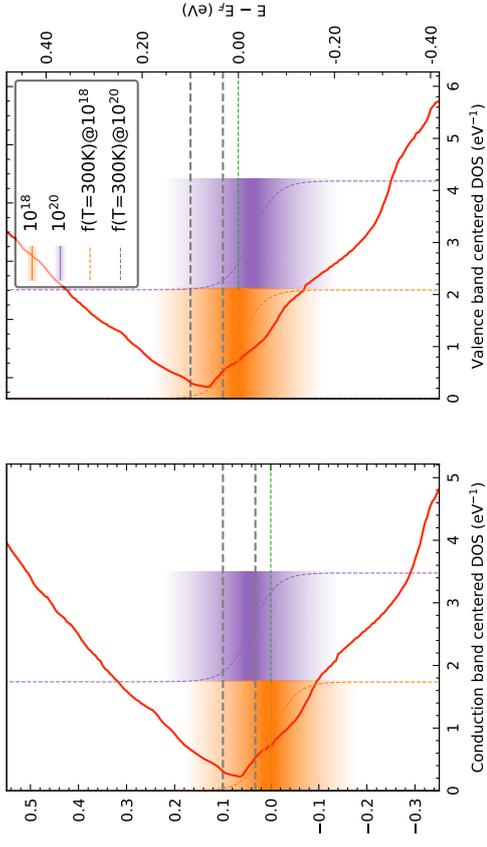

Conduction band centered DOS (eV$^{-1}$)

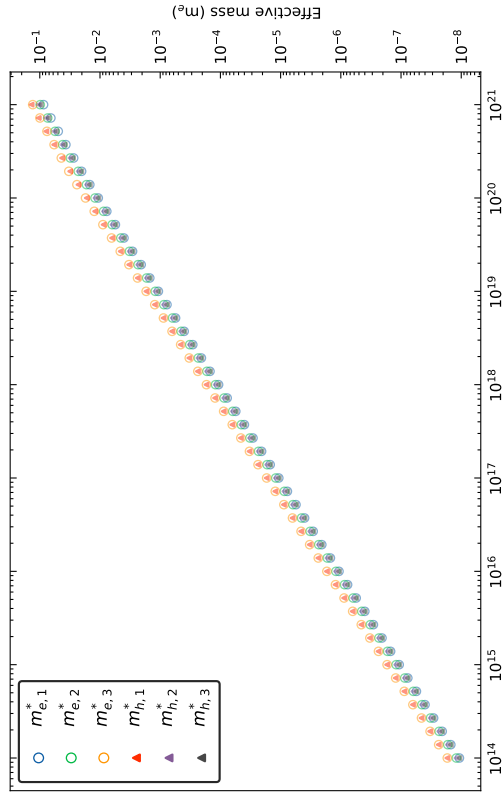

Valence band centered DOS (eV$^{-1}$)

Legend: $10^{18}$, $10^{20}$, f(T=300K)@$10^{18}$, f(T=300K)@$10^{20}$

Conduction- and valence-band centered DOS; orange(purple) shaded rectangles indicate energy windows used for doping concentration at n/p = $10^{18}$($10^{20}$) cm$^{-3}$; orange(purple) dashed lines show the corresponding Fermi energy positions for these dopings, and the gray dashed line for $E_F$ of zero doping.

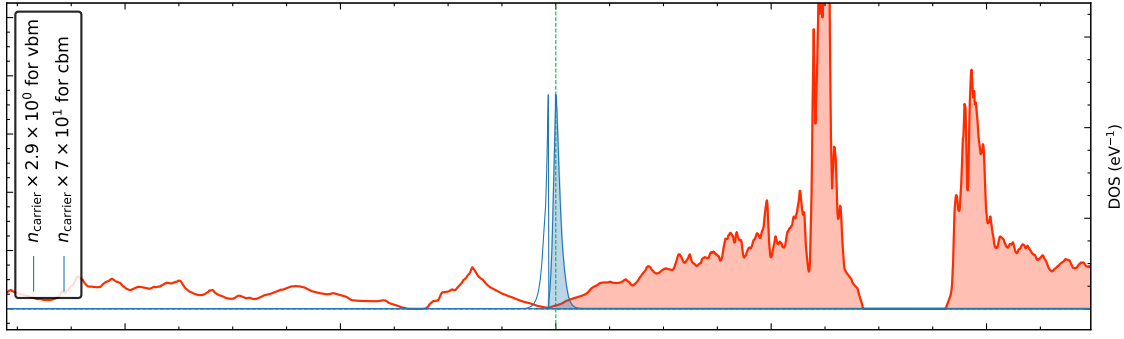

Effective masses vs doping in log-log scale. Markers: circles = electron effective mass $m^*_{e,i}$, triangles = hole effective mass $m^*_{h,i}$, where $i = 1, 2, 3$ denotes the three directions of eigenvectors of the effective mass tensor.

Legend: $m^*_{e,1}$, $m^*_{e,2}$, $m^*_{e,3}$, $m^*_{h,1}$, $m^*_{h,2}$, $m^*_{h,3}$

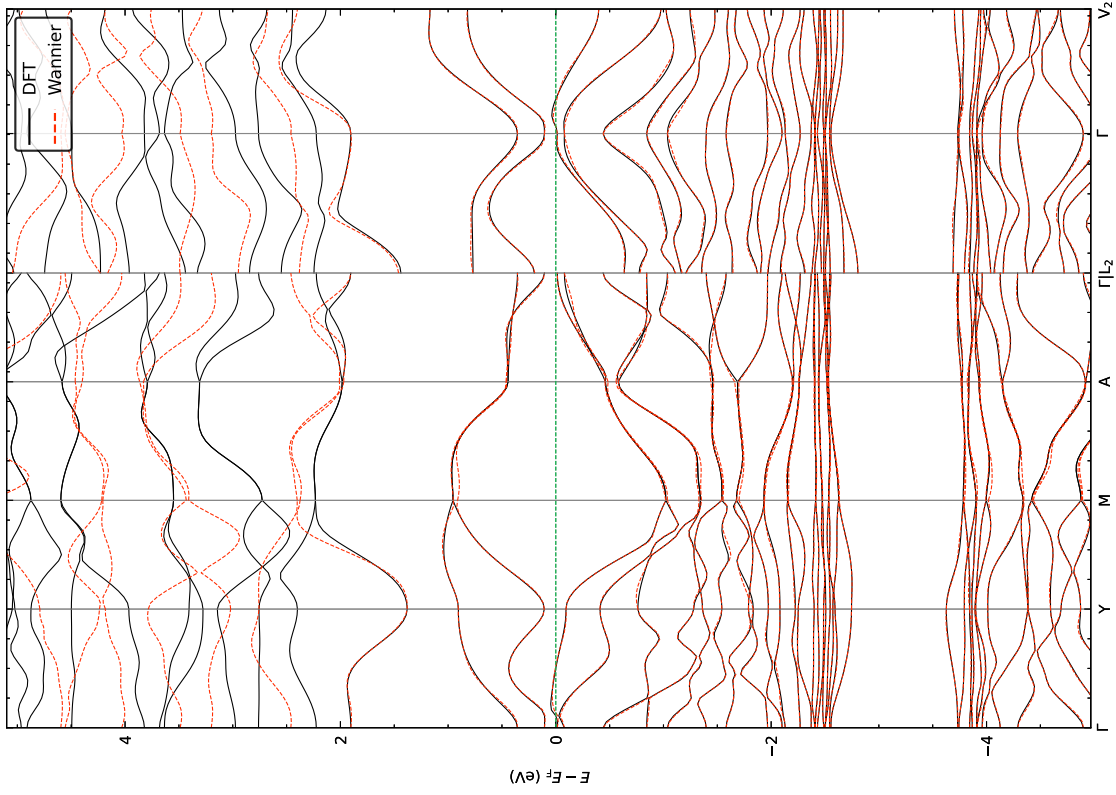

Band structure comparison: DFT (black solid) and Wannier (red dashed) with Fermi energy $E_F$ shifted to 0 eV.

Legend: $n_{carrier} \times 2.9 \times 10^0$ for vbm; $n_{carrier} \times 7 \times 10^1$ for cbm

Total density of states (DOS) of Wannier bands and number of carriers. Filled area shows occupied states.

# SiRh

| Formula IUPAC | SiRh | | Num. atoms | 8 |
|---|---|---|---|---|
| Formula Hill | Rh₄Si₄ | | Space group | P2₁/c (14) |
| Boltzwann UUID | 1fb5d50e-aaf5-4a4f4-a5fd-e6ce0ba0d244 | | Source database | MC3D ID |
| W90optwic UUID | d9fd8fd9-997f-4a76-ba4c-35a13c8t75e1b | | | mpds-S1301056 |
| Hole stability | 10000.00% | | Bandgap | 0.27 eV |
| Electron stability | 10000.00% | | Band distance | 3.4 meV |

## Eigenvalues of the effective mass tensors in electron mass units [$m_e$].
Absolute value shown for three doping levels: $10^{18}$, $10^{19}$, $10^{20}$ for electrons and holes.

| Effective masses | Doping $10^{18}$ | Doping $10^{19}$ | Doping $10^{20}$ |
|---|---|---|---|
| electron | (0.00052, **0.00086**, 0.00085) | (0.00052, **0.00086**, 0.00085) | (0.05, **0.09**, 0.08) |
| hole | (0.00052, **0.00086**, 0.00085) | (0.00052, **0.00086**, 0.00085) | (0.054, 0.082, **0.09**) |

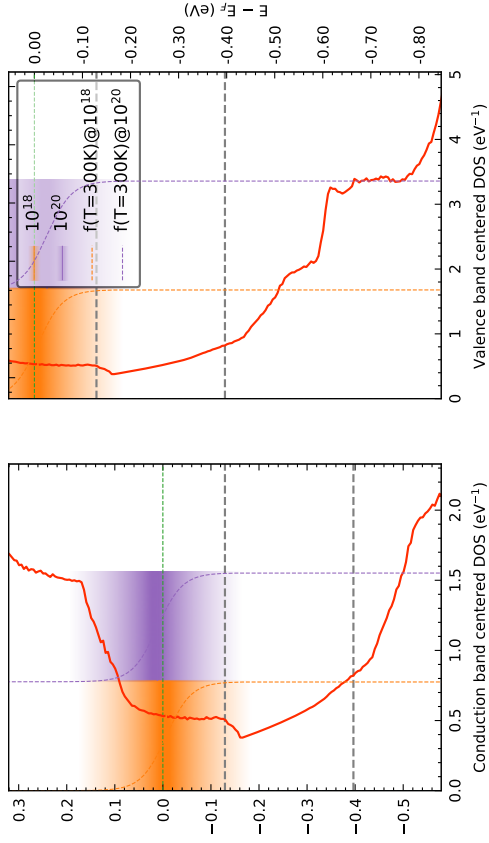

Conduction band centered DOS (eV⁻¹)

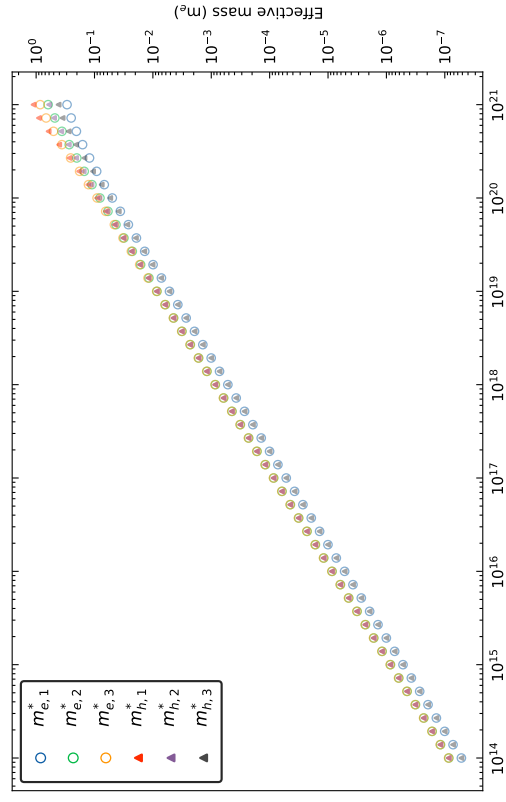

Valence band centered DOS (eV⁻¹)

Conduction- and valence-band centered DOS; orange(purple) shaded rectangles indicate energy windows used for doping concentration at n/p = $10^{18}$ ($10^{20}$) cm⁻³; orange(purple) dashed lines show the corresponding Fermi energy positions for these dopings, and the gray dashed line for $E_f$ of zero doping.

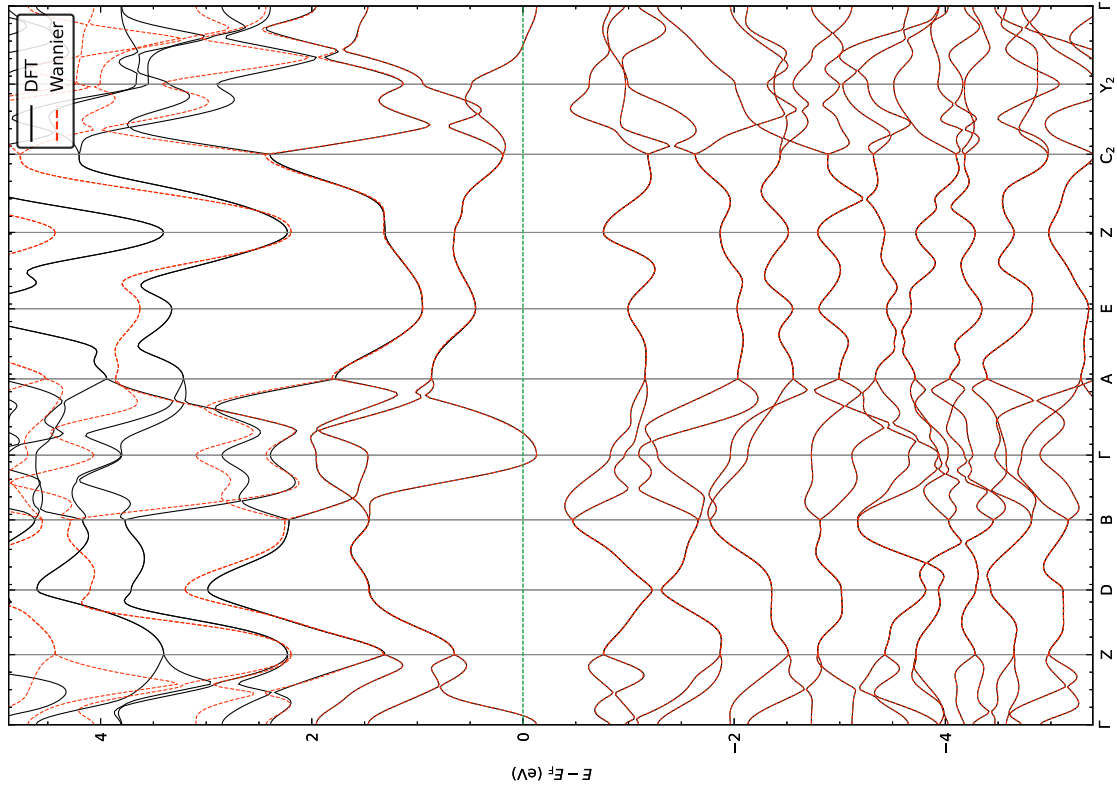

Effective masses vs doping in log-log scale. Markers: circles - electron effective mass $m^*_{e,i}$, triangles - hole effective mass $m^*_{h,i}$, where i = 1,2,3 denotes the three directions of eigenvectors of the effective mass tensor.

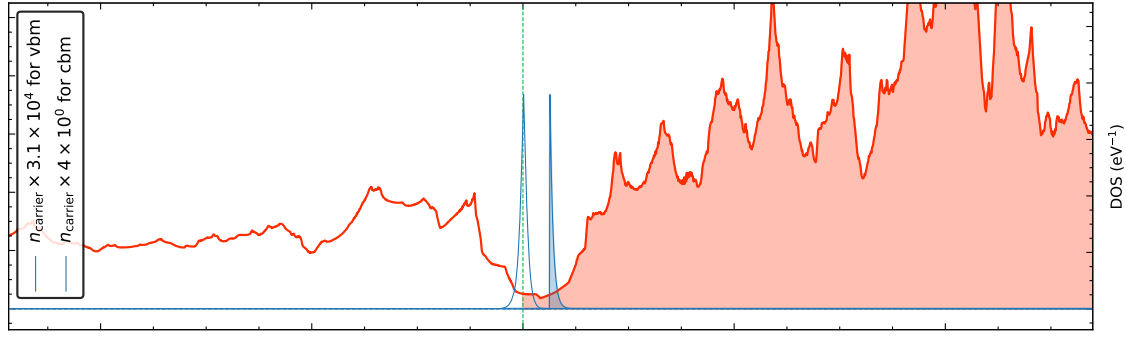

Total density of states (DOS) of Wannier bands and number of carriers. Filled area shows occupied states.

Band structure comparison: DFT (black solid) and Wannier (red dashed) with Fermi energy $E_f$ shifted to 0 eV.

# InAgS₂

| Formula IUPAC | InAgS₂ |
| --- | --- |
| Formula Hill | AgInS₂ |
| Boltzmann UUID | f8482b0f-cf0c-4b3a-8b08-c8b8f8ed1ed5c |
| W90optwkc UUID | 4b5c48b5b-c42c-4777-89eb-ec89047b0d60a |
| Electron stability | 9360.00% |

| Num. atoms | 4 |
| --- | --- |
| Space group | R-3m (166) |
| Source database | mp2d-41262/pbesol-v1 |
| MC3D ID | mpds-S1010630 |
| Bandgap | 0.10 eV |
| Hole stability | 10000.00% |
| Band distance | 3.6 meV |

**Eigenvalues of the effective mass tensors in electron units [$m_e$].**
Absolute value shown for three doping levels: $10^{18}$, $10^{19}$, $10^{20}$ for electrons and holes.

| Effective masses | Doping $10^{18}$ | | | Doping $10^{19}$ | | | Doping $10^{20}$ | | |
| --- | --- | --- | --- | --- | --- | --- | --- | --- | --- |
| electron | (0.00096, | 0.00096, | 0.0005) | (0.0006, | 0.0096, | 0.0005) | (0.009, | 0.088, | 0.07) |
| hole | (0.00096, | 0.00096, | 0.0005) | (0.00096, | 0.00096, | 0.0005) | (0.0006, | 0.050, | 0.09) |

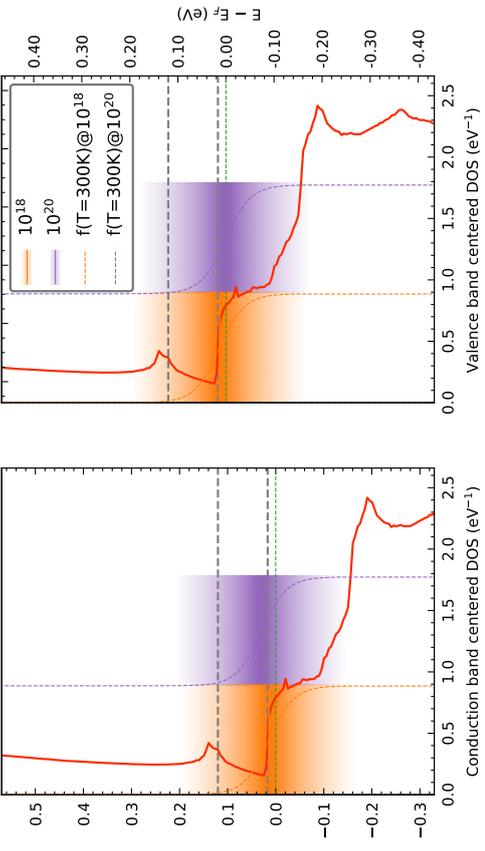

Conduction band centered DOS (eV⁻¹)

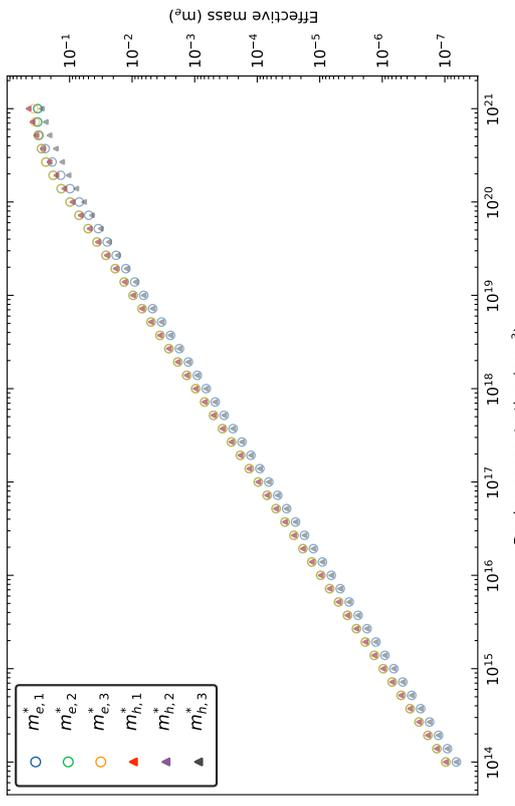

Valence band centered DOS (eV⁻¹)

Conduction- and valence-band centered DOS. orange(purple) shaded rectangles indicate energy windows used for doping concentration at n/p = $10^{18}$ ($10^{20}$) cm⁻³; orange(purple) dashed lines show the corresponding Fermi energy positions for these dopings, and the gray dashed line for $E_f$ of zero doping.

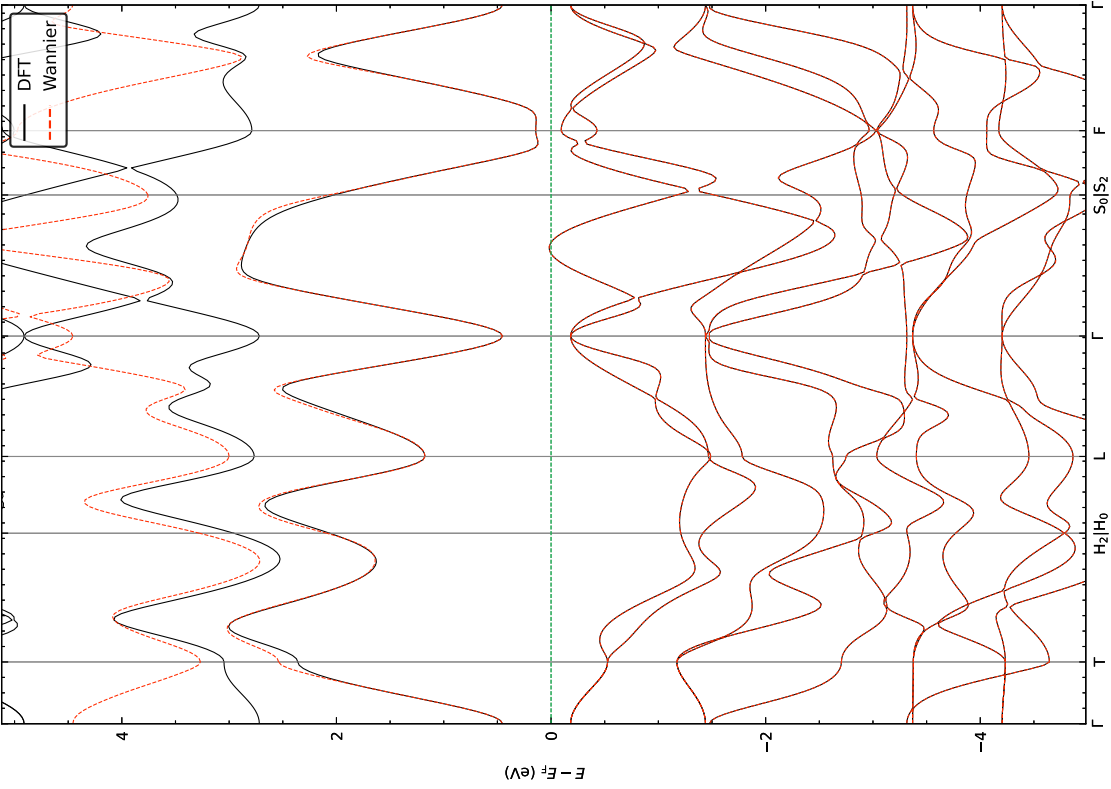

Effective masses vs doping in log-log scale. Markers: circles = electron effective mass $m_{e,i}^*$, triangles = hole effective mass $m_{h,i}^*$, where i = 1,2,3 denotes the three directions of eigenvectors of the effective mass tensor.

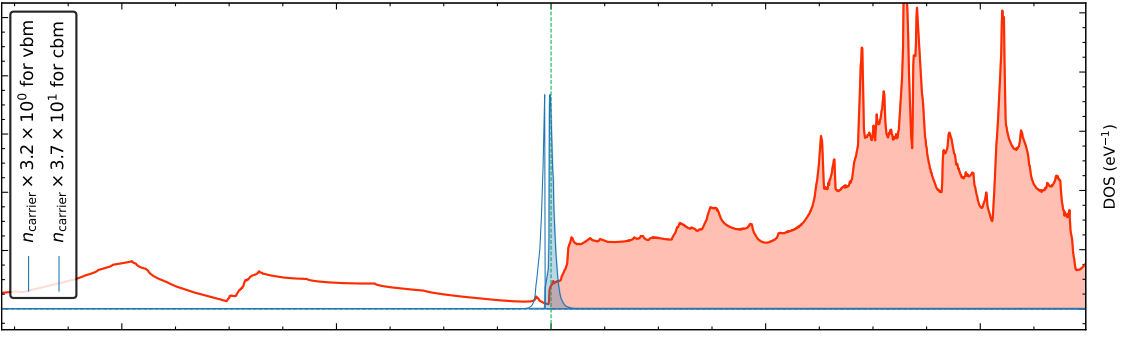

Total density of states (DOS) of Wannier bands and number of carriers. Filled area shows occupied states.

$n_{carrier} \times 3.2 \times 10^6$ for vbm
$n_{carrier} \times 3.7 \times 10^4$ for cbm

DOS (eV⁻¹)

Band structure comparison: DFT (black solid) and Wannier (red dashed) with Fermi energy $E_f$ shifted to 0 eV.

## SnS

| Formula IUPAC | SnS |
|---|---|
| Formula Hill | SSn |
| Space group | Fm-3m (225) |
| Boltzmann UUID | e4acd972-6969-4a88-a383-7c64d6d4796e |
| W90optwfac UUID | 2d6414a1-824d-428f-bc96-ab46a5f2fb09 |
| Source database | MC3D ID | nc3d-74977/pbesol-v1 |
| Bandgap | mp1s-S534209 |
| Hole stability | 39000.00% |
| Electron stability | 39000.00% |
| Num. atoms | 2 |
| Bandgap | 0.06 eV |
| Band distance | 6.9 meV |

**Eigenvalues of the effective mass tensors in electron mass units [$m_e$].**
Absolute value shown for three doping levels: $10^{18}$, $10^{19}$, $10^{20}$ for electrons and holes.

| Effective masses | | Doping $10^{18}$ | Doping $10^{19}$ | Doping $10^{20}$ |
|---|---|---|---|---|
| electron | ($m^*_{e,1}$, $m^*_{e,2}$, $m^*_{e,3}$) | (0.0021,0.0021,0.0021) | (0.019,0.019,0.019) | (0.076,0.076,0.076) |
| hole | ($m^*_{h,1}$, $m^*_{h,2}$, $m^*_{h,3}$) | (0.0021,0.0021,0.0021) | (0.02,0.02,0.02) | (0.083,0.083,0.083) |

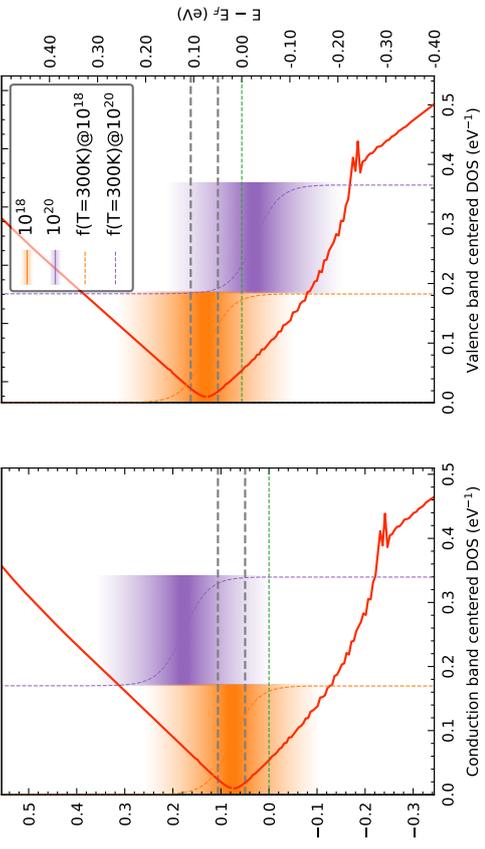
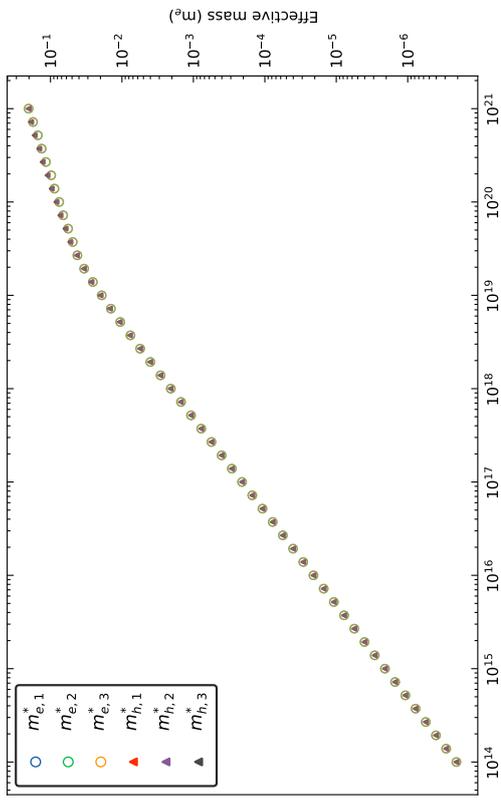

Conduction- and valence-band centered DOS: orange(purple) shaded rectangles indicate conduction- and valence-band centered energy windows used for doping concentration at n/p = $10^{18}$($10^{20}$) $cm^{-3}$, orange(purple) dashed lines show the corresponding Fermi energy positions for these dopings, and the gray dashed line for $E_F$ of zero doping.

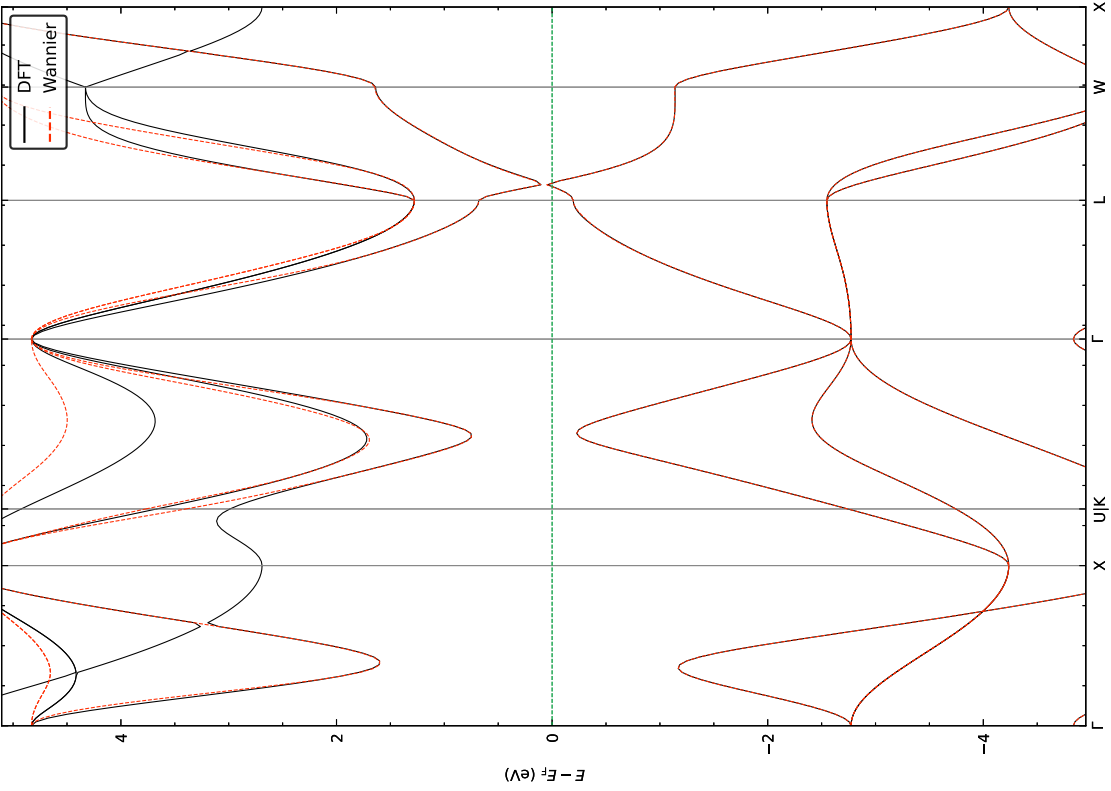

Effective masses vs doping in log-log scale. Markers: circles = electron effective mass $m^*_{e,i}$, triangles = hole effective mass $m^*_{h,i}$, where i = 1,2,3 denotes the three directions of eigenvectors of the effective mass tensor.

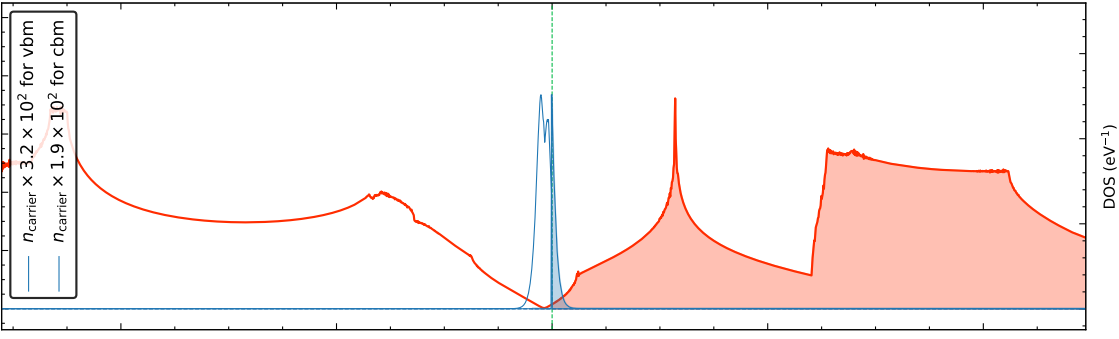

Total density of states (DOS) of Wannier bands and number of carriers. Filled area shows occupied states.

$n_{carrier} \times 3.2 \times 10^2$ for vbm
$n_{carrier} \times 1.9 \times 10^2$ for cbm

Band structure comparison: DFT (black solid) and Wannier (red dashed) with Fermi energy $E_F$ shifted to 0 eV.

12

# Rb₂Hg₃Se₄

| | | | |
|---|---|---|---|
| Formula IUPAC | Rb₂Hg₃Se₄ | Num. atoms | 18 |
| Formula Hill | Hg₃Rb₂Se₄ | Space group | Ibam (72) |
| Boltzwann UUID | 6d11424c-f209-40ce-bb0a-7196413b6a5 | Source database | mc3d-4116/phesol-v1 |
| WBBoptwisc UUID | 58aa3d9b-b11c-4e7c-9ad4-c67c965bccec | MC3D ID | icsd-430729 |
| Electron stability | 21000.00% | Bandgap | 1.03 eV |
| Hole stability | 1600.00% | Band distance | 3.9 meV |

**Eigenvalues of the effective mass tensors in electron mass units [$m_e$].**
Absolute value shown for three doping levels: $10^{18}$, $10^{19}$, $10^{20}$ for electrons and holes.

| Effective masses | Doping $10^{18}$ | Doping $10^{19}$ | Doping $10^{20}$ |
|---|---|---|---|
| electron | (0.024, **0.0053**, 0.0025) | (0.0091, **0.0075**, 0.0069) | (0.055, **0.046**, 0.043) |
| hole | (0.0031, **0.0062**, 0.0032) | (0.039, **0.056**, 0.043) | (0.43, **0.88**, 0.69) |

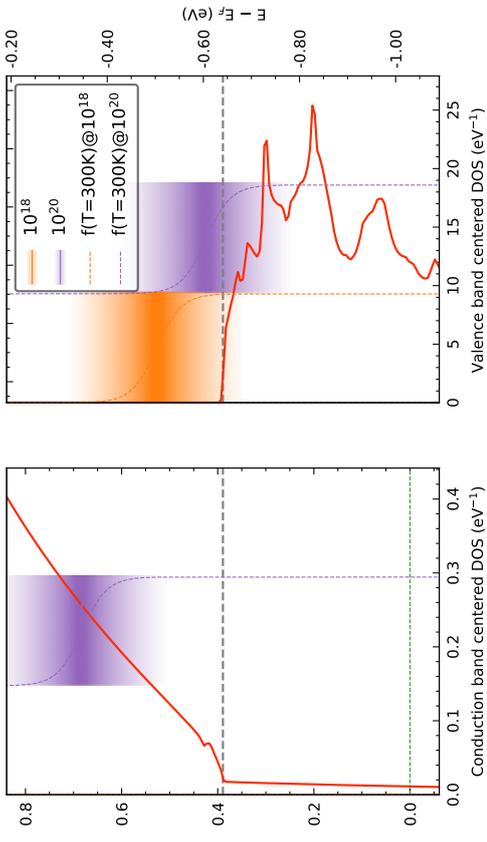

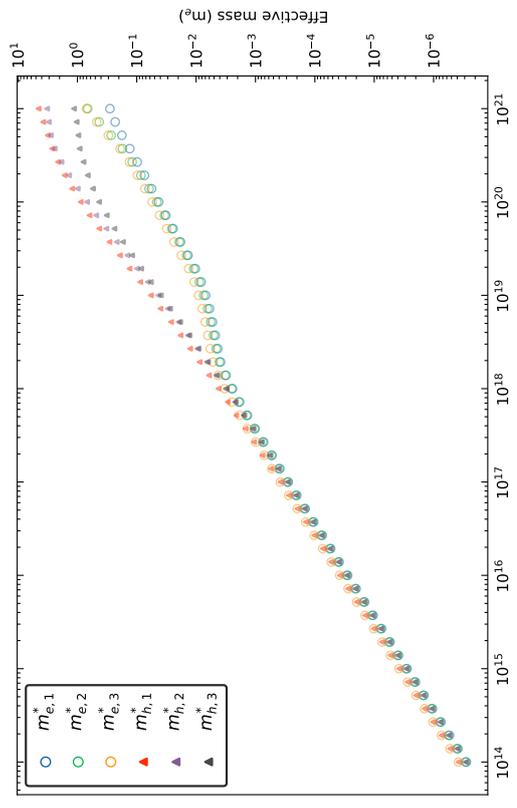

Conduction- and valence-band centered DOS; orange(purple) shaded rectangles indicate energy windows used for doping concentration at $n/p = 10^{18}$ ($10^{20}$) $cm^{-3}$; orange(purple) dashed lines show the corresponding Fermi energy positions for these dopings, and the gray dashed line for $E_F$ of zero doping.

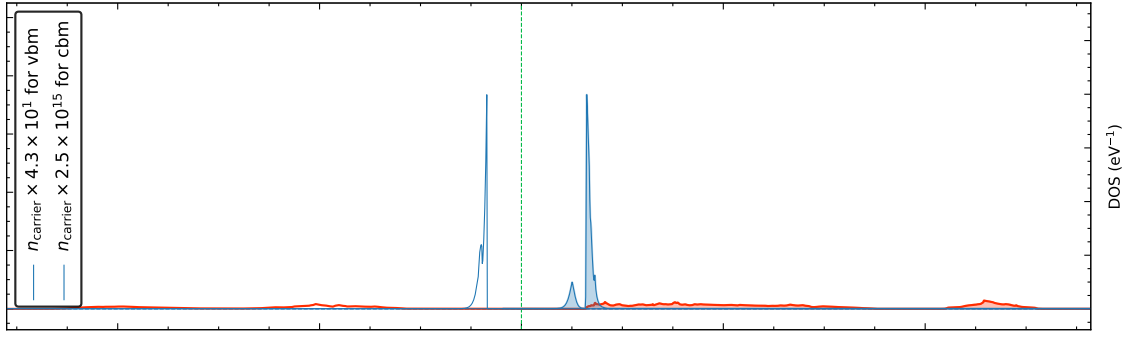

Band structure comparison: DFT (black solid) and Wannier (red dashed) with Fermi energy $E_F$ shifted to 0 eV.

Total density of states (DOS) of Wannier bands and number of carriers. Filled area shows occupied states.

Legend:
$n_{carrier} \times 4.3 \times 10^1$ for vbm
$n_{carrier} \times 2.5 \times 10^{15}$ for cbm

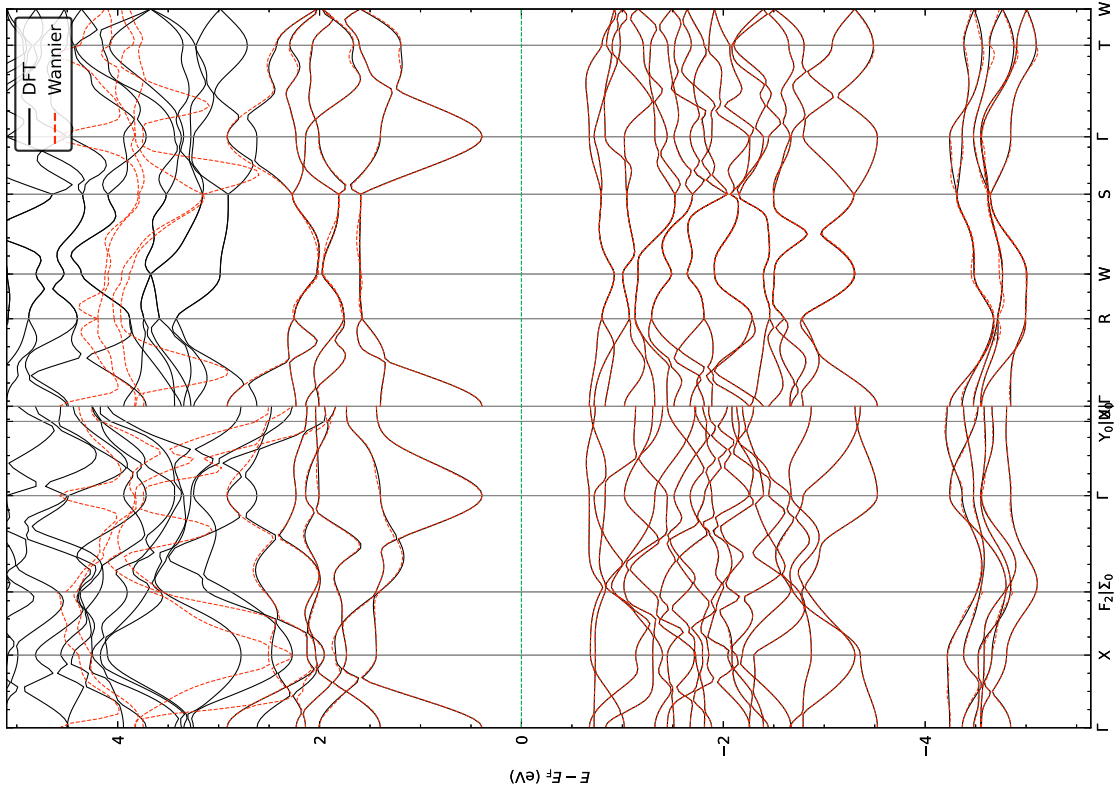

Effective masses vs doping in log-log scale. Markers: circles = electron effective mass $m^*_{e,i}$, triangles = hole effective mass $m^*_{h,i}$, where $i = 1, 2, 3$ denotes the three directions of eigenvectors of the effective mass tensor.

Legend: $m^*_{e,1}$, $m^*_{e,2}$, $m^*_{e,3}$, $m^*_{h,1}$, $m^*_{h,2}$, $m^*_{h,3}$

# K$_2$(NbCl$_3$)$_3$

| | |
|---|---|
| Formula IUPAC | K$_2$(NbCl$_3$)$_3$ |
| Formula Hill | Cl$_{18}$K$_6$Nb$_6$ |
| Space group | C2/m (12) |
| Source database | MC3D |
| MC3D ID | mc3d-24544/pbesol-v1 |
| Wannier UUID | 9222fd00-c889-4fcd-89f7-ae8088b0f1d4 |
| W90optpwke UUID | 07ae6f63c-a2c5-4934-a12a-0fe614e7b38f |
| Boltzwann UUID | nmpids-s31702971 |
| Bandgap | 1.14 eV |
| Band distance | 3.7 meV |
| Num. atoms | 28 |
| Hole stability | 4300.00% |
| Electron stability | 6000.00% |

**Eigenvalues of the effective mass tensors in electron mass units [$m_e$].**

| | Doping 10$^{18}$ | Doping 10$^{19}$ | Doping 10$^{20}$ |
|---|---|---|---|
| Absolute value shown for three doping levels: 10$^{18}$, 10$^{19}$, 10$^{20}$ for electrons and holes. | | | |
| electron | (0.0035, 0.003, 0.0017) | (0.023, 0.0017, 0.0099) | (0.22, 0.17, 0.096) |
| hole | (0.0027, 0.0022, 0.0013) | (0.013, 0.011, 0.0064) | (0.12, 0.086, 0.057) |

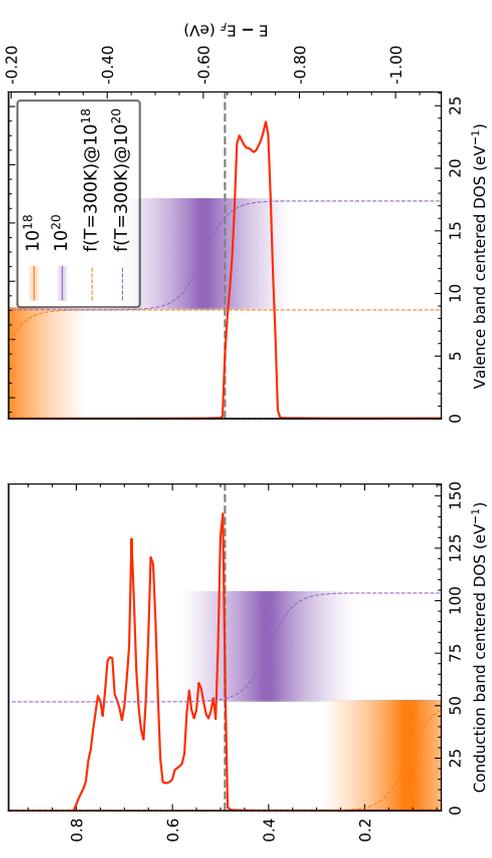

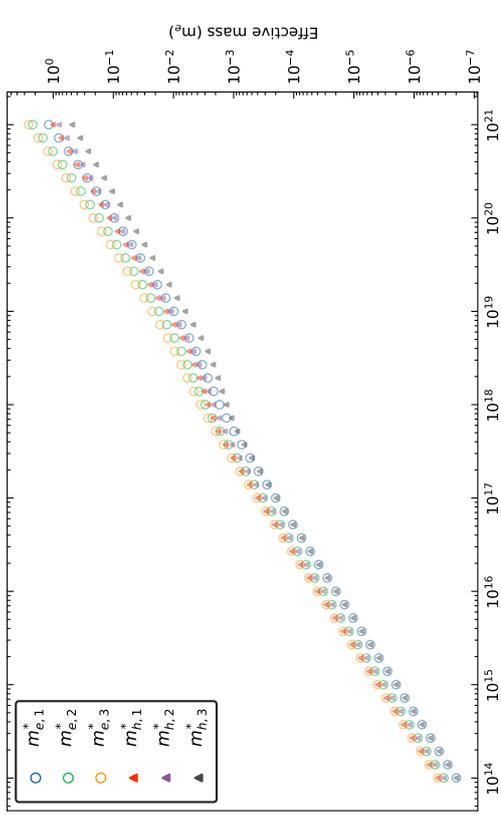

Conduction- and valence-band centered DOS. orange(purple) shaded rectangles indicate energy windows used for doping concentration at n/p = 10$^{18}$(10$^{20}$) cm$^{-3}$; orange(purple) dashed lines show the corresponding Fermi energy positions for these dopings, and the gray dashed line for $E_F$ of zero doping.

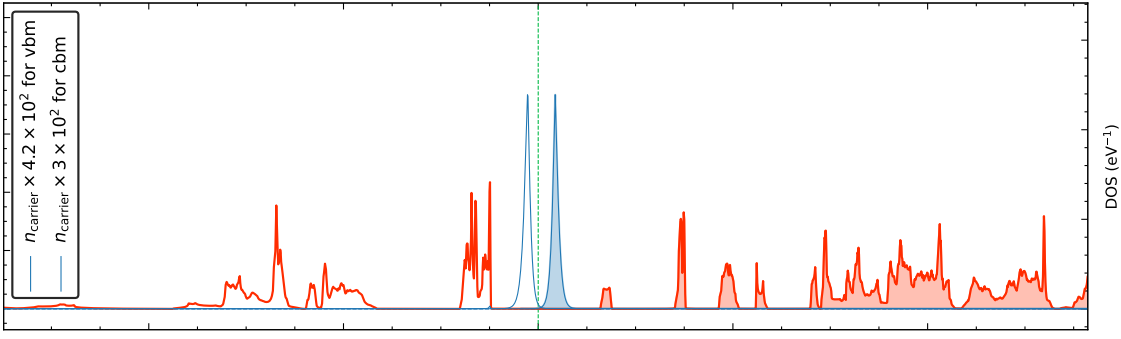

Total density of states (DOS) of Wannier bands and number of carriers. Filled area shows occupied states.

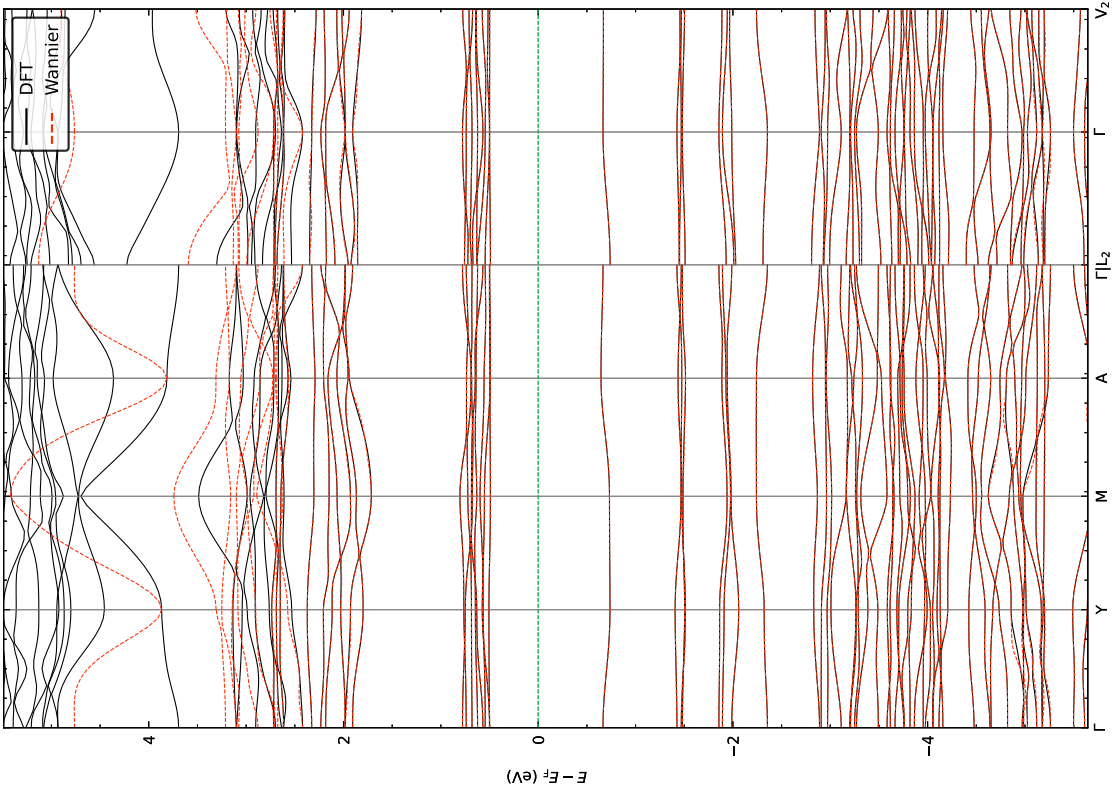

Band structure comparison: DFT (black solid) and Wannier (red dashed) with Fermi energy $E_F$ shifted to 0 eV.

Effective masses vs doping in log-log scale. Markers: circles = electron effective mass $m_{e,i}^*$, triangles = hole effective mass $m_{h,i}^*$, where $i$ = 1, 2, 3 denotes the three directions of eigenvectors of the effective mass tensor.

| Formula IUPAC | GeSe |
| Formula Hill | GeSe |
| W90optwic UUID | b159fafc-9bc6-4da3-b445-576 e75f2l8b1 |
| Boltzwann UUID | 35d641d5-0696-484b-8fd3-9601a4abcd90 |
| Electron stability | **1800.00%** |

| Num. atoms | 2 |
| Space group | Fm-3m (225) |
| Source database | MC3D 1D |
| Boltzwann UUID | mc3d-28887/pbesol-v1 |
| Bandgap | mpds-S1631l219 |
| Hole stability | 0.17 eV |
| | **1800.00%** |
| Band distance | 4.1 meV |

**Eigenvalues of the effective mass tensors in electron mass units [$m_e$].**
Absolute value shown for three doping levels: $10^{18}$, $10^{19}$, $10^{20}$ for electrons and holes.

| Effective masses | Doping $10^{18}$ | Doping $10^{19}$ | Doping $10^{20}$ |
|---|---|---|---|
| electron | (0.0061, 0.0061, 0.0061) | (0.047, 0.047, 0.047) | (0.051, 0.051, 0.051) |
| hole | (0.0061, 0.0061, 0.0061) | (0.12, 0.12, 0.12) | (0.12, 0.12, 0.12) |

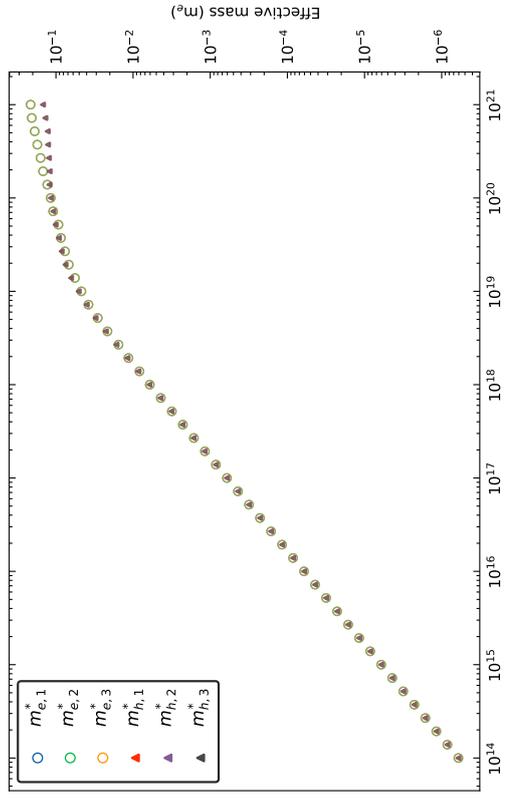

Conduction- and valence-band centered DOS; orange(purple) shaded rectangles indicate energy windows used for doping concentration at n/p = $10^{18}$ ($10^{20}$) $cm^{-3}$; orange(purple) dashed lines show the corresponding Fermi energy positions for these dopings, and the gray dashed line for $E_f$ of zero doping.

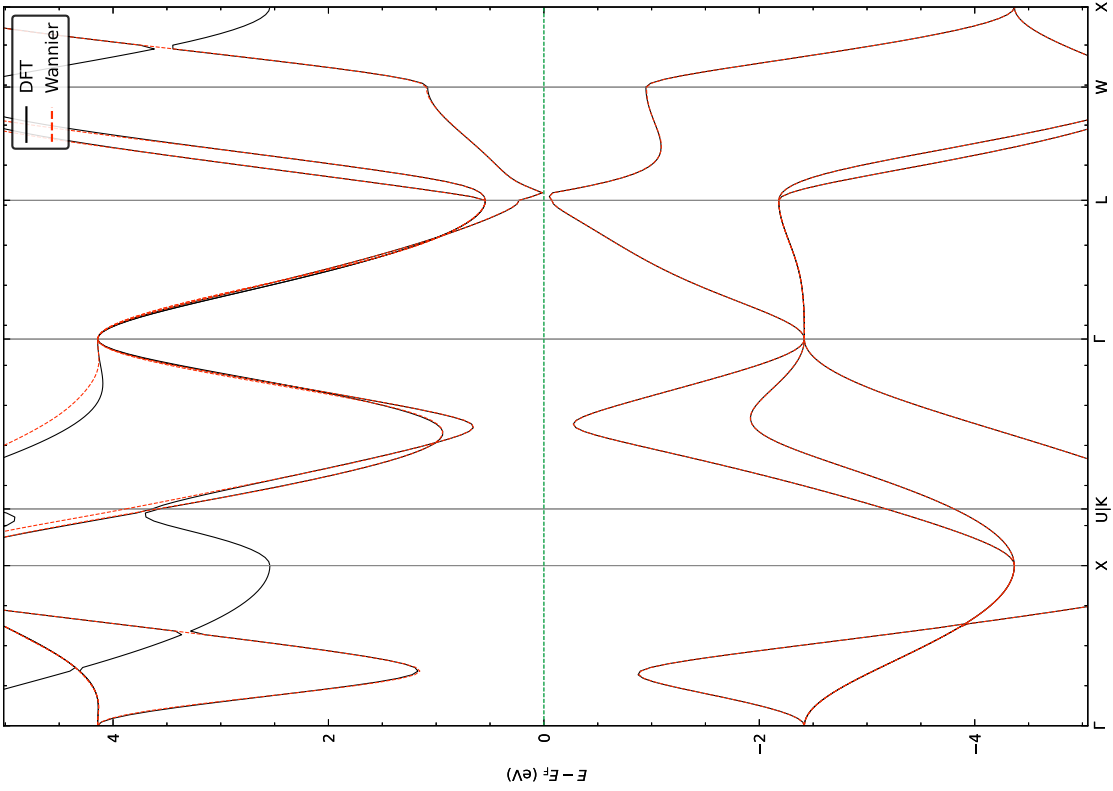

Effective masses vs doping in log-log scale. Markers: circles = electron effective mass $m^*_{e,i}$, triangles = hole effective mass $m^*_{h,i}$, where $i = 1,2,3$ denotes the three directions of eigenvectors of the effective mass tensor.

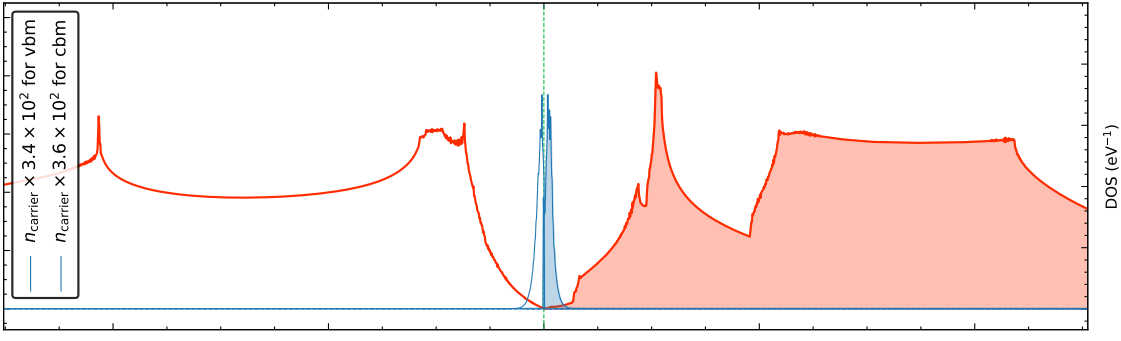

Total density of states (DOS) of Wannier bands and number of carriers. Filled area shows occupied states.

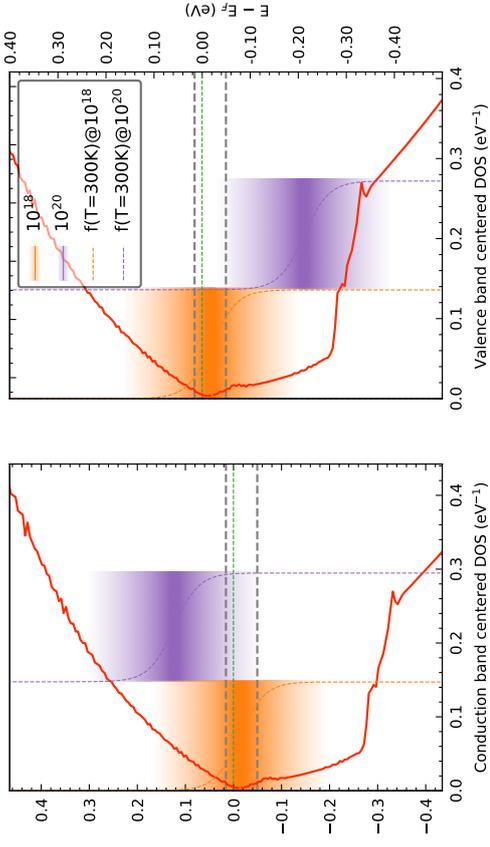

Band structure comparison: DFT (black solid) and Wannier (red dashed) with Fermi energy $E_f$ shifted to 0 eV.

# FeSi

| Formula IUPAC | FeSi | Num. atoms | 8 |
|---|---|---|---|
| Formula Hill | Fe8Si8 | Space group | P2_13 (198) |
| Boltzwann UUID | 9c08f5b5-a2c7-4c64-97fc-a242a0221e43 | MC3D ID | mc3d-12152/pbesol-v1 |
| W90optwckc UUID | 1d050edc-289f-4b8a-8d94-81d8a6946b4 | Source database | mpds-S1023808 |
| Hole stability | 10000.00% | Bandgap | 0.15 eV |
| Electron stability | 9300.00% | Band distance | 0.80 meV |

**Eigenvalues of the effective mass tensors in electron mass units [$m_e$].**
Absolute value shown for three doping levels: $10^{18}$, $10^{19}$, $10^{20}$ cm$^{-3}$ for electrons and holes.

| Effective masses | Doping $10^{18}$ | Doping $10^{19}$ | Doping $10^{20}$ |
|---|---|---|---|
| electron | (**0.009**, 0.009, 0.069) | (**0.009**, 0.009, 0.069) | (0.65, 0.65, **0.05**) |
| hole | (**0.009**, 0.009, 0.069) | (**0.009**, 0.009, 0.069) | (**0.071**, 0.71, 0.71) |

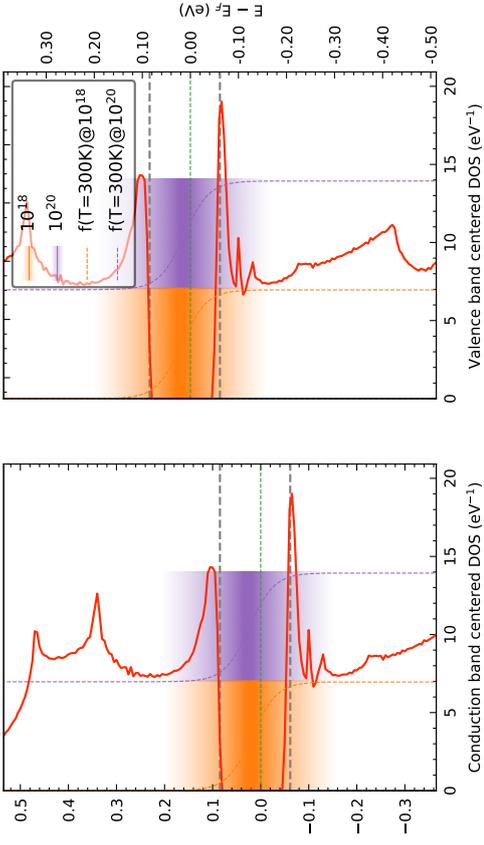

Conduction band centered DOS (eV$^{-1}$)

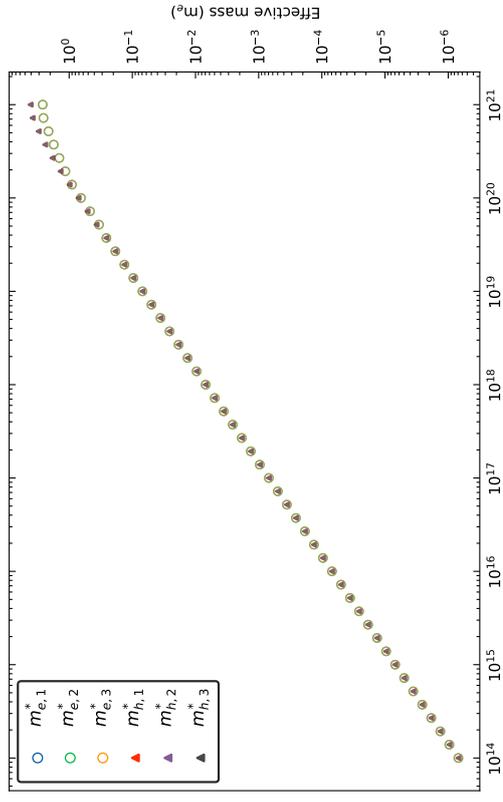

Valence band centered DOS (eV$^{-1}$)

Conduction- and valence-band centered DOS: orange(purple) shaded rectangles indicate energy windows used for doping concentration at n/p = $10^{18}$($10^{20}$) cm$^{-3}$; orange(purple) dashed lines show the corresponding Fermi energy positions for these dopings, and the gray dashed line for $E_F$ of zero doping.

Inset labels: $10^{18}$, $10^{20}$, f(T=300K)@$10^{18}$, f(T=300K)@$10^{20}$

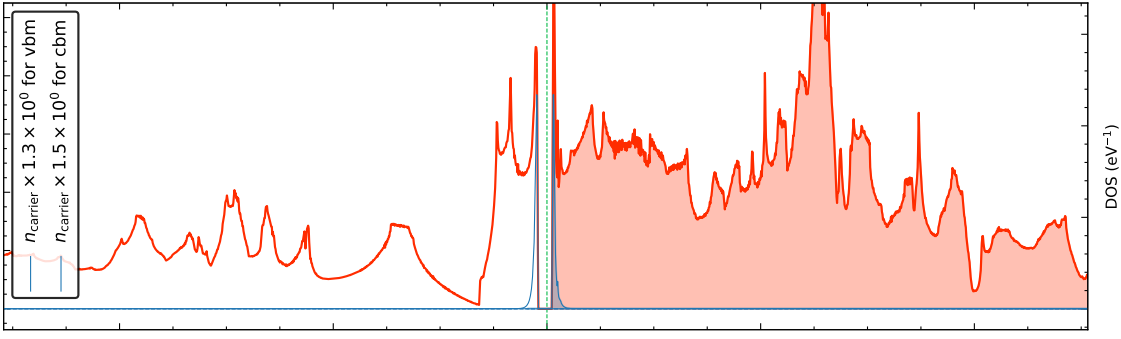

DOS (eV$^{-1}$)

Legend: $n_{carrier} \times 1.3 \times 10^6$ for vbm; $n_{carrier} \times 1.5 \times 10^5$ for cbm

Total density of states (DOS) of Wannier bands and number of carriers. Filled area shows occupied states.

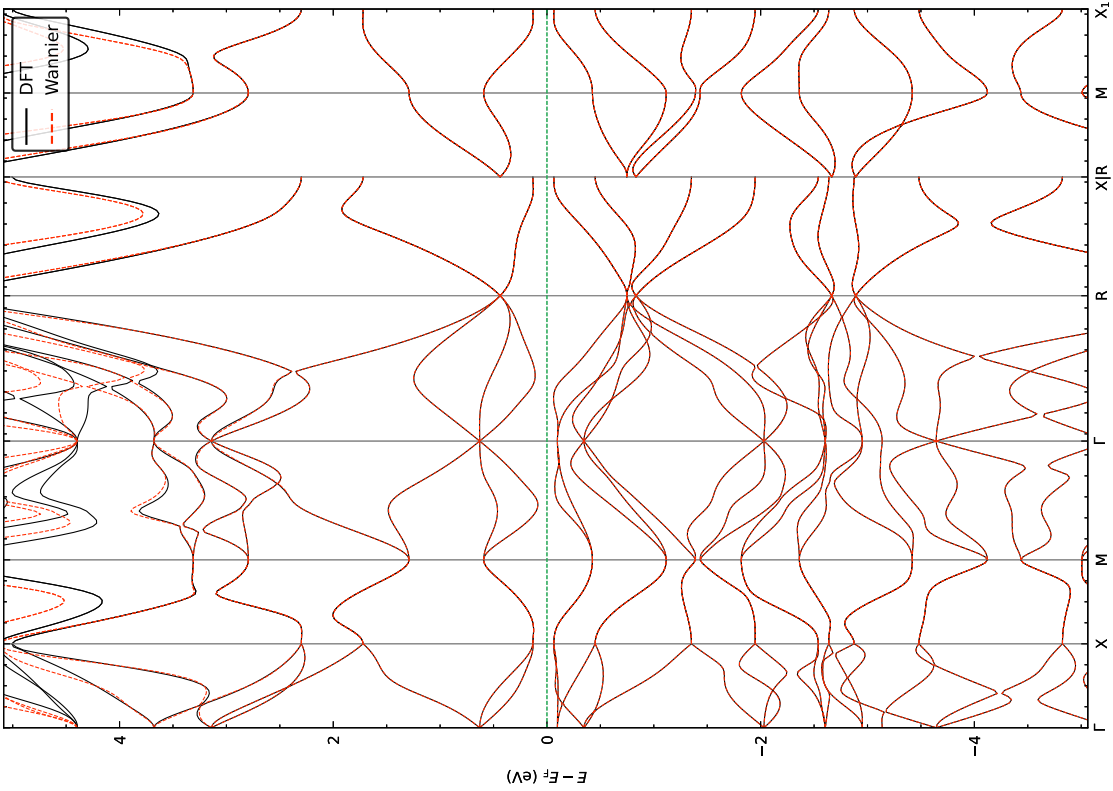

Band structure comparison: DFT (black solid) and Wannier (red dashed) with Fermi energy $E_F$ shifted to 0 eV.

Legend: DFT, Wannier

Effective masses vs doping in log-log scale. Markers: circles = electron effective mass $m^*_{e,i}$, triangles = hole effective mass $m^*_{h,i}$, where i = 1,2,3 denotes the three directions of eigenvectors of the effective mass tensor.

Legend: $m^*_{e,1}$, $m^*_{e,2}$, $m^*_{e,3}$, $m^*_{h,1}$, $m^*_{h,2}$, $m^*_{h,3}$

# AgO

| Formula IUPAC | AgO | | Num. atoms | 8 |
|---|---|---|---|---|
| Formula Hill | Ag₄O₄ | | Space group | P2₁/c (14) |
| Boltzwann UUID | d0423f0a-f51d-4b4f-947c-92f9f0c29799 | | MC3D ID | mc3d-5517/pbesol-v1 |
| W90optwkc UUID | 59149b0e9-527b-4a6b-b4f6-851e2a49ba51 | | Source database | mpds-S1251008 |
| Hole stability | 83b0.b0% | | Bandgap | 0.08 eV |
| Electron stability | 83b0.b0% | | Band distance | 0.38 meV |

**Eigenvalues of the effective mass tensors in electron mass units [m_e].**
Absolute value shown for three doping levels: $10^{18}$, $10^{19}$, $10^{20}$ for electrons and holes.

| Effective masses | | Doping $10^{18}$ | Doping $10^{19}$ | Doping $10^{20}$ |
|---|---|---|---|---|
| electron | | (0.0069, 0.0028, 0.0067) | (0.009, 0.028, 0.067) | (0.64, 0.18, **0.66**) |
| hole | | (0.0069, 0.0028, 0.0067) | (0.008, 0.031, 0.064) | (**0.58**, 0.41, 0.53) |

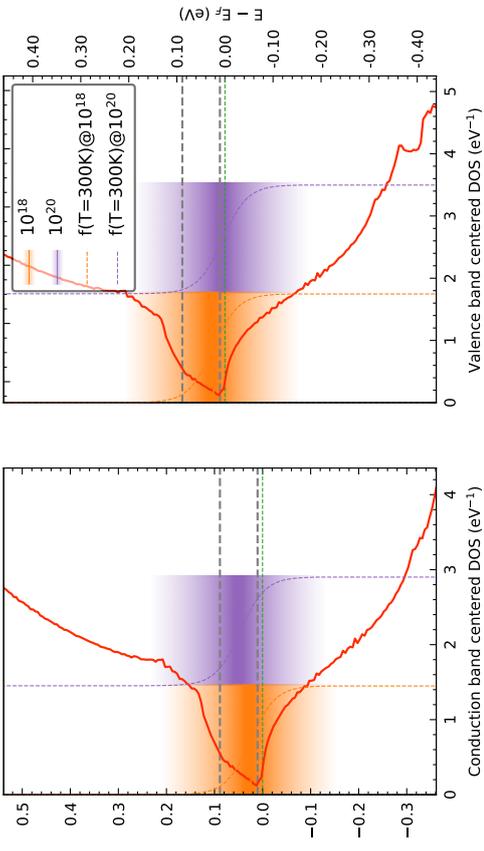

Conduction band centered DOS (eV⁻¹)

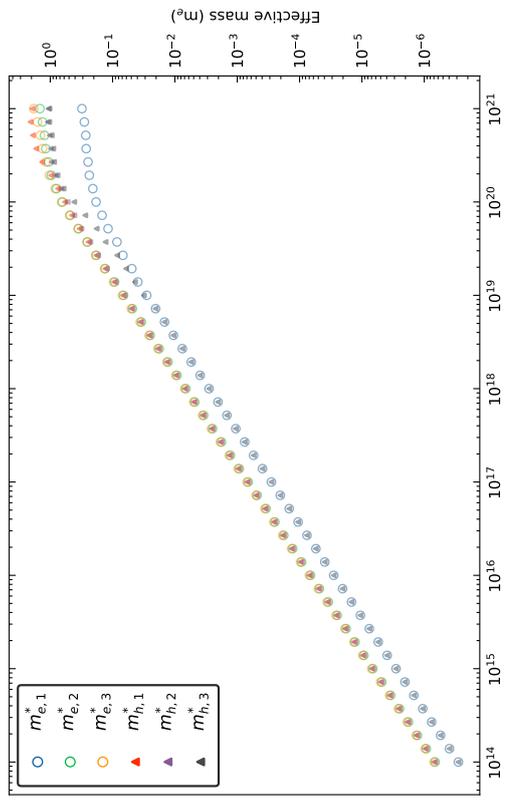

Valence band centered DOS (eV⁻¹)

$10^{18}$
$10^{20}$
f(T=300K)@$10^{18}$
f(T=300K)@$10^{20}$

Conduction- and valence-band centered DOS; orange(purple) shaded rectangles indicate energy windows used for doping concentration at n/p = $10^{18}$($10^{20}$) cm⁻³; orange(purple) dashed lines show the corresponding Fermi energy positions for these dopings, and the gray dashed line for $E_F$ at zero doping.

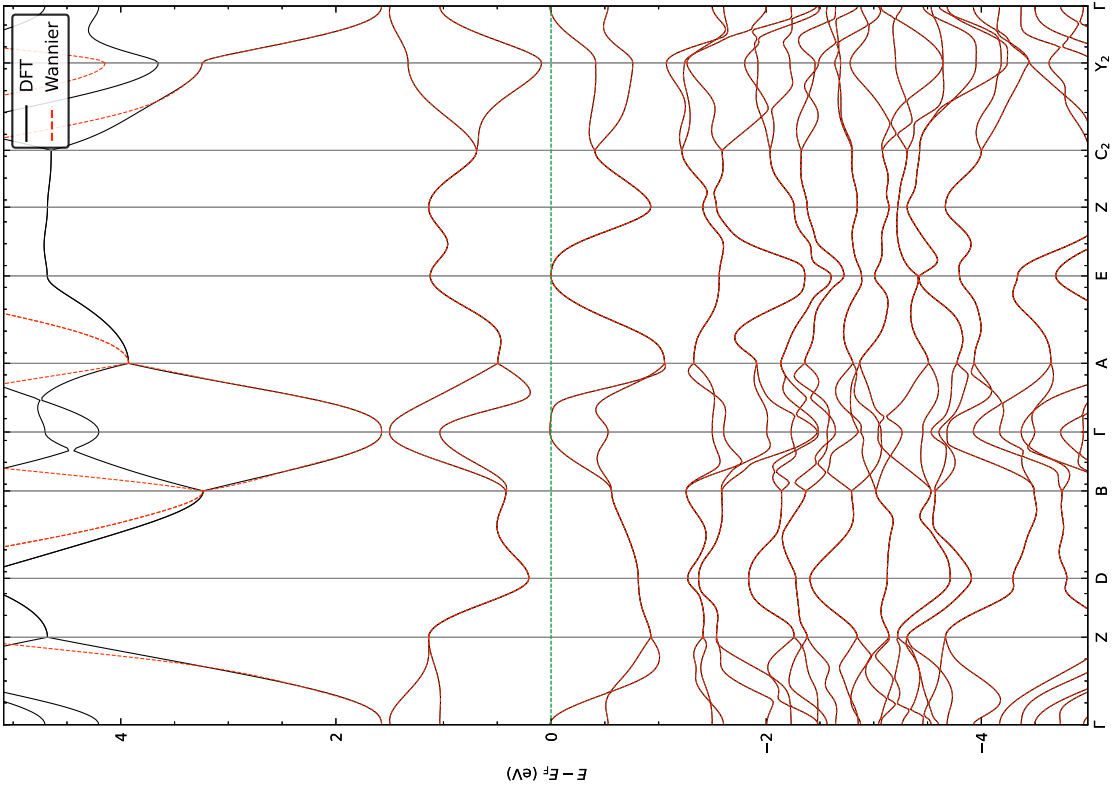

Effective masses vs doping in log-log scale. Markers: circles = electron effective mass $m^*_{e,i}$, triangles = hole effective mass $m^*_{h,i}$, where i = 1,2,3 denotes the three directions of eigenvectors of the effective mass tensor.

$m^*_{e,1}$, $m^*_{e,2}$, $m^*_{e,3}$, $m^*_{h,1}$, $m^*_{h,2}$, $m^*_{h,3}$

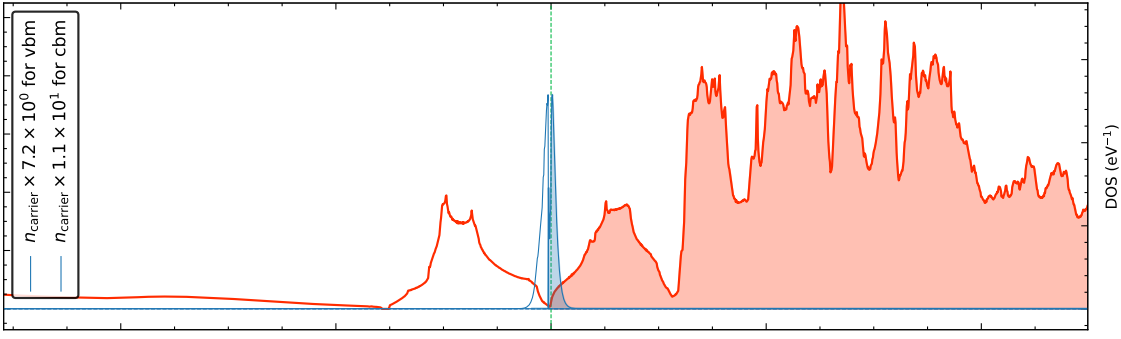

$n_{carrier} \times 7.2 \times 10^6$ for vbm
$n_{carrier} \times 1.1 \times 10^5$ for cbm

DOS (eV⁻¹)

Total density of states (DOS) of Wannier bands and number of carriers. Filled area shows occupied states.

DFT
Wannier

Band structure comparison: DFT (black solid) and Wannier (red dashed) with Fermi energy $E_F$ shifted to 0 eV.

17

# SnTe

| | | | |
|---|---|---|---|
| Formula IUPAC | SnTe | Num. atoms | 2 |
| Formula Hill | SnTe | Space group | Fm-3m (225) |
| MC3D ID | mc3d-55818 / phesol-v1 | Source database | mpds-S532444 |
| Wyckoff UUID | 008f6c2d-bf5b-4a7d-831b-4a61e8f71b6f | Bandgap | 0.06 eV |
| Boltzmann UUID | bae69ed7-fa5d-48e7-ab5a-28f59e75ce09 | Band distance | 5.7 meV |
| Hole stability | 1300.00% | | |
| Electron stability | 1700.00% | | |

**Eigenvalues of the effective mass tensors in electron mass units [$m_e$].**
Absolute value shown for three doping levels $10^{18}$, $10^{19}$, $10^{20}$ $cm^{-3}$ for electrons and holes.

| Effective masses | Doping $10^{18}$ | Doping $10^{19}$ | Doping $10^{20}$ |
|---|---|---|---|
| electron | 0.007, 0.007, 0.007 | 0.053, 0.053, 0.053 | 0.13, 0.13, 0.13 |
| | (0.0071, 0.0071, 0.0071) | (0.053, 0.053, 0.053) | (0.13, 0.13, 0.13) |
| hole | 0.0071, 0.0071, 0.0071 | 0.053, 0.053, 0.053 | 0.096, 0.096, 0.096 |
| | (0.0071, 0.0071, 0.0071) | (0.053, 0.053, 0.053) | (0.096, 0.096, 0.096) |

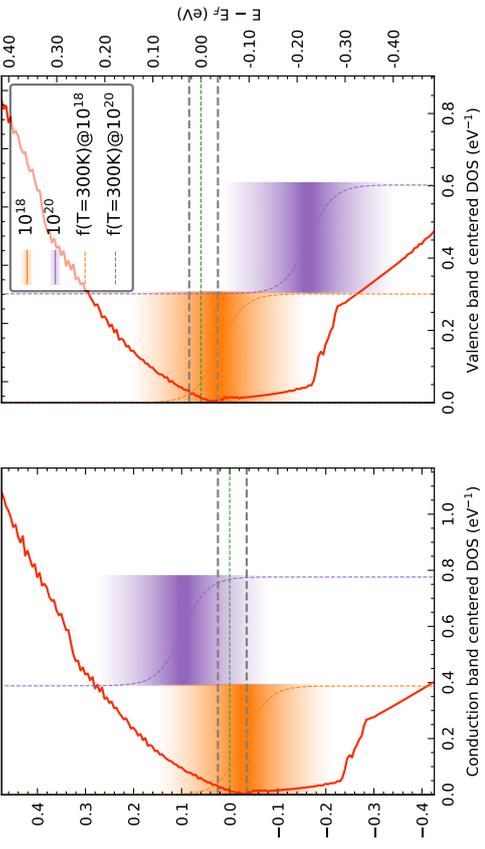

Conduction band centered DOS (eV$^{-1}$)

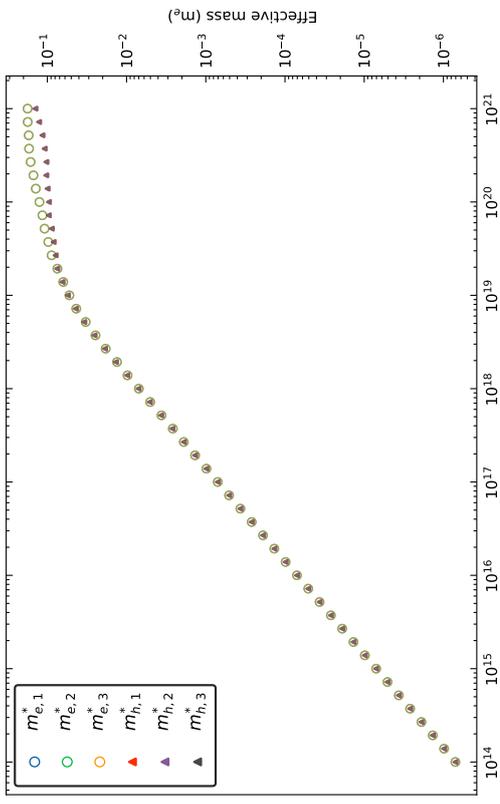

Valence band centered DOS (eV$^{-1}$)

Conduction- and valence-band centered DOS: orange(purple) shaded rectangles indicate energy windows used for doping concentration at n/p = $10^{18}$ ($10^{20}$) $cm^{-3}$; orange(purple) dashed lines show the corresponding Fermi energy positions for these dopings, and the gray dashed line for $E_F$ of zero doping.

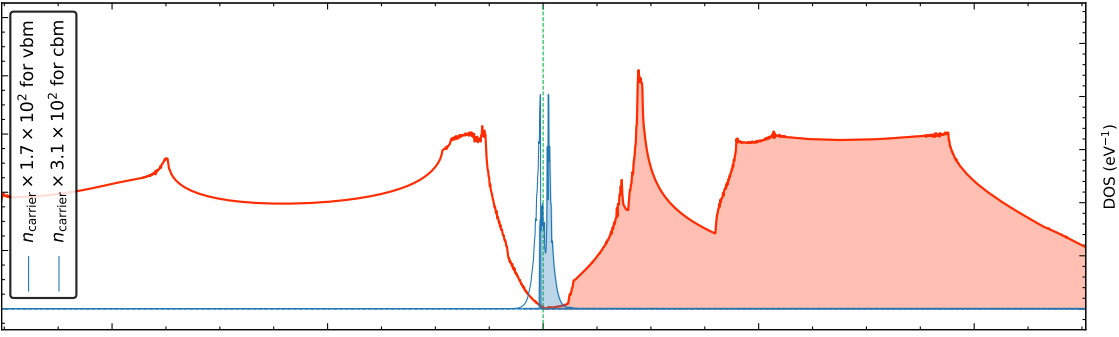

Band structure comparison: DFT (black solid) and Wannier (red dashed) with Fermi energy $E_F$ shifted to 0 eV.

Total density of states (DOS) of Wannier bands and number of carriers. Filled area shows occupied states.

$n_{carrier} \times 1.7 \times 10^{2}$ for vbm
$n_{carrier} \times 3.1 \times 10^{2}$ for cbm

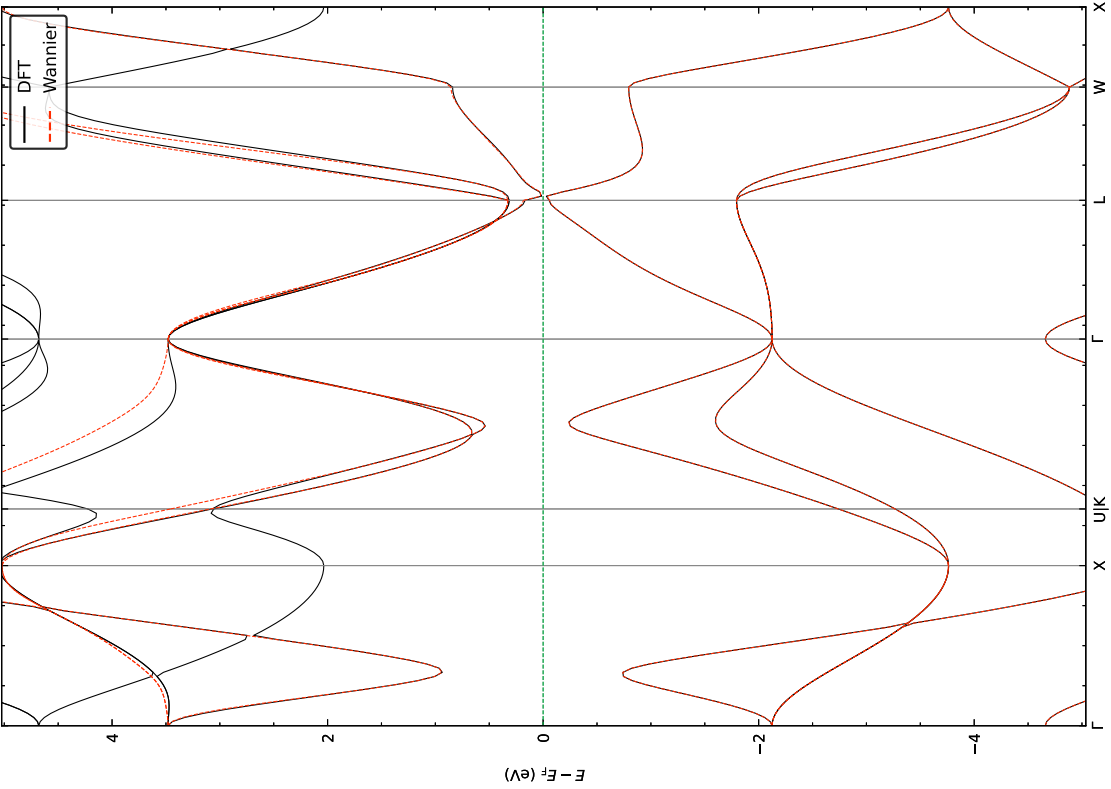

Effective masses vs doping in log-log scale. Markers: circles = electron effective mass $m^*_{e,i}$, triangles = hole effective mass $m^*_{h,i}$, where i = 1,2,3 denotes the three directions of eigenvectors of the effective mass tensor.

Legend: $m^*_{e,1}$, $m^*_{e,2}$, $m^*_{e,3}$, $m^*_{h,1}$, $m^*_{h,2}$, $m^*_{h,3}$



\end{document}